\RequirePackage{fix-cm}

\documentclass[smallcondensed]{svjour3}     
\smartqed  
\usepackage[utf8]{inputenc} 
\usepackage[T1]{fontenc}    
\usepackage{hyperref}       
\usepackage{url}            
\usepackage{booktabs}       
\usepackage{amsfonts}       
\usepackage{nicefrac}       
\usepackage{microtype}      

\usepackage{blindtext}
\usepackage{textcomp}
\usepackage{graphicx}
\usepackage{float}
\usepackage[section]{placeins}
\usepackage[position=bottom,aboveskip=2pt]{subcaption}
\usepackage[nottoc]{tocbibind}
\usepackage{siunitx}
\usepackage{layouts}
\usepackage{enumitem, amssymb, amsmath}
\usepackage{bbm}
\usepackage{xcolor}
\usepackage[export]{adjustbox}
\usepackage[comma,authoryear,round]{natbib}

\journalname{Arxiv}
\begin{document}

\title{TRU-NET: A Deep Learning Approach to High Resolution Prediction of Rainfall }

\titlerunning{TRU-NET}        

\author{
  Rilwan A.~Adewoyin \and Peter Dueben \and Peter Watson \and Yulan He \and Ritabrata Dutta
}
\institute{
  Rilwan A.~Adewoyin
  \at Department of Computer Science, University of Warwick, UK\\
Department of Computer Science and Engineering,\\ Southern University of Science and Technology, China
  \and
  Peter Dueben
  \at
  Earth System Modelling Section, The European Centre for Medium-Range Weather Forecasts
  \and
  Peter Watson
  \at 
  School of Geographical Sciences, University of Bristol, UK
  \and
  Yulan He
  \at
  Department of Computer Science, University of Warwick, UK
    \and
  Ritabrata Dutta
  \at Department of Statistics, University of Warwick, UK 
}

\authorrunning{Adewoyin et al.} 


\maketitle

\begin {abstract}

Climate models (CM) are used to evaluate the impact of climate change on the risk of floods and heavy precipitation events. However, these numerical simulators produce outputs with low spatial resolution that exhibit difficulties representing precipitation events accurately. This is mainly due to computational limitations on the spatial resolution used when simulating multi-scale weather dynamics in the atmosphere. 

To improve the prediction of high resolution precipitation we apply a Deep Learning (DL) approach using 
input data from a reanalysis product, that is comparable to a climate model's output, but can be directly related to precipitation observations at a given time and location. Further, our input excludes local precipitation, but includes model fields (weather variables) that are more predictable and generalizable than local precipitation. 

To this end, we present TRU-NET (Temporal Recurrent U-Net), an encoder-decoder model featuring a novel 2D cross attention mechanism between contiguous convolutional-recurrent layers to effectively model multi-scale spatio-temporal weather processes. We also propose a non-stochastic variant of the conditional-continuous (CC) loss function to capture the zero-skewed patterns of rainfall. Experiments show that our models, trained with our CC loss, consistently attain lower RMSE and MAE scores than a DL model prevalent in precipitation downscaling
and outperform a state-of-the-art dynamical weather model. Moreover, by evaluating the performance of our model under various data formulation strategies, for the training and test sets, we show that there is enough data for our deep learning approach to output robust, high-quality results across seasons and varying regions.


\end{abstract}

\section{Introduction and Background}
\label{sec:Introduction}
Across the globe, society is becoming increasingly prone to extreme precipitation events due to climate change. The United Nations stated that flooding was the most dominant weather-related disaster over the 20 years to 2015, affecting 2.3 billion people and accounting for \$1.89 trillion in reported economic losses \citep{2015_UNreport_thehumancostof}. With the increase in the monetary and societal risk posed by flooding \citep{IPCC2019Special_report}, the CM predictions for extreme precipitation events are an important resource in guiding the decision of policy-makers.

State-of-the-art regional climate models (RCM) typically run at horizontal resolutions of $\sim$2-25 km for simulations. These simulations provide an approach to get local detail, but they are computationally expensive and must be developed separately for each climate model (CM) \citep{May2004,IPCC2007Synth_chapter8}.

Quantile mapping is a well researched bias correction method \citep{https://doi.org/10.1029/2001JD000659, 10.1007/978-981-15-5077-5_55} which aims to learn a mapping from quantiles of the empirical distribution of a CM's rainfall predictions to quantiles of the corresponding observed rainfall distribution. Multivariate quantile mapping techniques \citep{2018ClDy...50...31C,10.1029/2011WR011524}, improve upon their univariate counterparts by leveraging the correlation between rainfall and other weather variables and learning a mapping to observed rainfall from the empirical multivariate distribution of a CM's predictions for several weather variables. 
However, since quantile mapping methods use precipitation as an input we can expect their performance to weaken when using input data that is not from the CM during training. This is because precipitation simulation is a major challenge in climate modelling, and each climate model's precipitation simulation contains a unique precipitation bias profile. These bias profiles includes factors such as the degrees of: overestimation of number of days with rain; underestimation of extreme rainfall events; or prematurity of simulated seasonal weather patterns related to rain.

In contrast, our approach performs downscaling while only utilising input weather variables that are generally less sensitive to the choice of climate model used to generate them. As such, our approach is expected to be more robust to the climate model it is trained on.

The \textbf{main aim} of this paper is to create a model that can produce high resolution predictions for daily rainfall across the UK. During training, the input to our model is the low resolution model field data, excluding rainfall, from a reanalysis product. Precipitation observations for the related time and location are used as the target. This reanalysis input data is comparable to the output of a climate model and includes weather variables which, unlike precipitation, are well simulated by climate models. When trained, our model can then be used on the output of any CM simulation for the future, allowing us to produce computationally cheap long-term forecasts of high resolution precipitation into the future, helping to diagnose changes in precipitation events due to climate change. 

Our task can be interpreted as a combination of the two DL tasks of Image Super-Resolution and Sequence Transduction. DeepSD \citep{Vandal_2018} extended a popular Image Super-Resolution model to downscale 2D precipitation images up to a factor of 4x. They also incorporated an optimization scheme utilising a probabilistic conditional-continuous (CC) loss function for improving the modelling of
zero-skewed precipitation events. In the related sequence transduction task, precipitation nowcasting, \citep{ConvLSTM_shi} introduced the Convolutional Long Short-Term Memory (ConvLSTM) cell to simultaneously model the behaviour of weather dynamics in space and time, using convolutions to incorporate the surrounding flow-fields and a recurrent structure to model the temporal dependencies of weather. Extending upon this, the encoder-decoder ConvLSTM \citep{ConvLSTM_shi} is also able to represent weather dynamics defined on multiple scales in \textbf{space} due to the use of successive convolution based layers, each layer to model larger scale dynamics.

The two deep learning approaches we propose are the hierarchical Convolutional Gated Recurrent Unit model (HCGRU) and TRU-NET. Our TRU-NET model extends an encoder-decoder ConvLSTM by adding the ability to represent weather dynamics in multiple scales in \textbf{time} and includes the following three features:
\begin{enumerate}
\item A novel Fused Temporal Cross Attention (FTCA) mechanism to improve upon existing methods \citep{hierachicalLSTM_Lastcell,hierachical_LSTM_lasthiddenstate2,HierachicalLSTM_selfattn} to model multiple temporal scales of weather dynamics using a stacked recurrent structure. 

\item A encoder-decoder structure adapting the U-NET \citep{ronneberger2015unet} structure, by contracting or expanding the temporal dimensions as opposed to the spatial dimensions. 

\item A non probabilistic adaptation of the conditional continuous \citep{Husak2007_GammadistrtoreprRain,Vandal_2018} as a training loss objective which improves the prediction of extreme precipitation events, by decoupling the modelling of low intensity precipitation events and high intensity precipitation events.

\end{enumerate}

We present HCGRU, TRU-NET and the conditional continuous loss in Section~\ref{sec:models} and discuss model training details in Section~\ref{sec:training}. In Section~\ref{sec:experiment}, we perform experiments to compare TRU-NET to baselines and investigate TRU-NET's performance on out-of-sample forecasting tasks. Finally, we perform an ablation study to compare our proposed FTCA against alternative existing methods. The results of these experiments show that: 

\begin{itemize}
    \item Our novel model, TRU-NET, achieves a lower RMSE and MAE than both a state-of-the-art dynamical weather forecasting model's coarse precipitation prediction and U-NET.  
    \item The quality of TRU-NET's predictions are stable when evaluated on out-of-sample weather predictions formed by time periods outside of the training set.
    \item Our proposed FTCA outperforms existing methods to decrease the temporal scale modelled by successive recurrent layers in a stacked recurrent structure.
\end{itemize}

\section{Data} 
\label{sec:data}
We use the ERA5 reanalysis dataset \citep{ERA5} as an analogue for CM output. Reanalysis data is based on a weather forecast model -- that is running at a similar resolution to a CM -- to which observations are constantly assimilated to yield the best estimate of the weather state. We take these historical weather state estimates and use it to predict the high resolution precipitation observations, made available by the E-OBS rainfall dataset \citep{E_obs}. The E-OBS rainfall dataset is created from the observational records of a network of weather stations. In order to get an estimate for precipitation at points on a structured geographic grid, E-obs uses a pre-trained spatial correlation function. Generally, E-Obs is a high quality dataset, having been found to have strong correlation with the highest quality UK precipitation dataset \citep{https://doi.org/10.1029/2009JD011799}. However, inspection revealed that the E-obs precipitation values had significant bias towards smaller rainfall values, especially under extreme precipitation events \citep{https://doi.org/10.1029/2009JD011799}.


\begin{figure*}[htbp]
\centering
    \includegraphics[width=\textwidth]{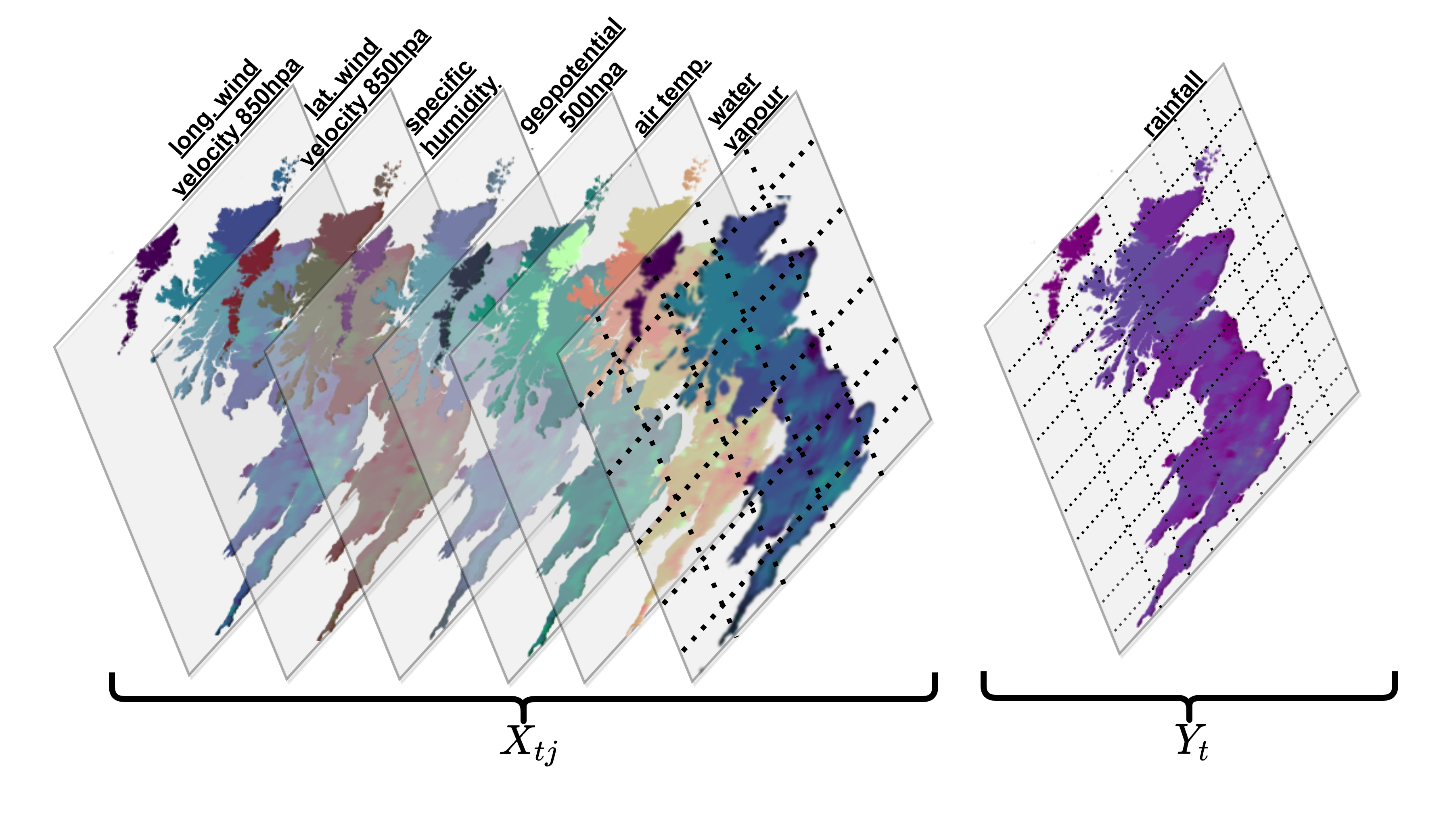}
  \caption{  \textbf{Illustration of Input to TRU-NET and HCGRU: } The input datum $X_{tj}, j \in [1\text{-}4] $ is obtained from ERA5 reanalysis dataset, while the target datum $Y_t$ is from the E-OBS dataset. An input datum is formed as a stack of 6 model fields including longitudinal and latitudinal components of wind velocity at 850 hPa, specific humidity, geopotential at 500hpa, air temperature and total column water vapour in the entire vertical column. These model fields are defined at 6-hourly resolution on a $20 {\times} 21$ grid over the United Kingdom. A target datum $Y_t$ is defined at a daily resolution on a $100 {\times} 140$ grid over the UK. The grids in the figure are not drawn to relative scale. }
\label{fig:model_input_output}
\end{figure*}

More concretely, our input data is formed as a timeseries of length $t$ days ($t\in[1,T]$) containing 6 key model fields (air temperature, specific humidity, longitudinal and latitudinal components of wind velocity at 850 hPa, geopotential height at 500 hPa and total column water vapour in the entire vertical column), each defined on a ($20 {\times} 21$) grid representing the UK at approximately 65km spatial resolution chosen to match that used in the UK Climate Projections datasets \citep{Murphy2018}. By stacking together the six model fields, as in Figure~\ref{fig:model_input_output}, we have a $(20, 21, 6)$ matrix, $X_{tj}, j\in[1\text{-}4]$, representing the UK weather state at 6-hour intervals. Our model will therefore take as input a sequence of daily model field observations $X_{t} \in \mathbb{R}^{4 {\times} 20 {\times} 21 {\times} 6}$, from which it will output a prediction for the true daily total precipitation ($mm$), $Y_t \in \mathbb{R}^{100 {\times} 140}$, defined on a $(100, 140)$ grid over the UK with approximately 8.5km spatial resolution and depicted in Figure~\ref{fig:model_input_output}.

\subsection{Experimental Settings}
\label{sec:DataExeperimentSettings}
\begin{figure}[htbp]
    \centering
    \begin{subfigure}[b]{0.47\linewidth}
          \includegraphics[angle=0,trim=0 0 75 550, clip, width=\linewidth,height=8cm]{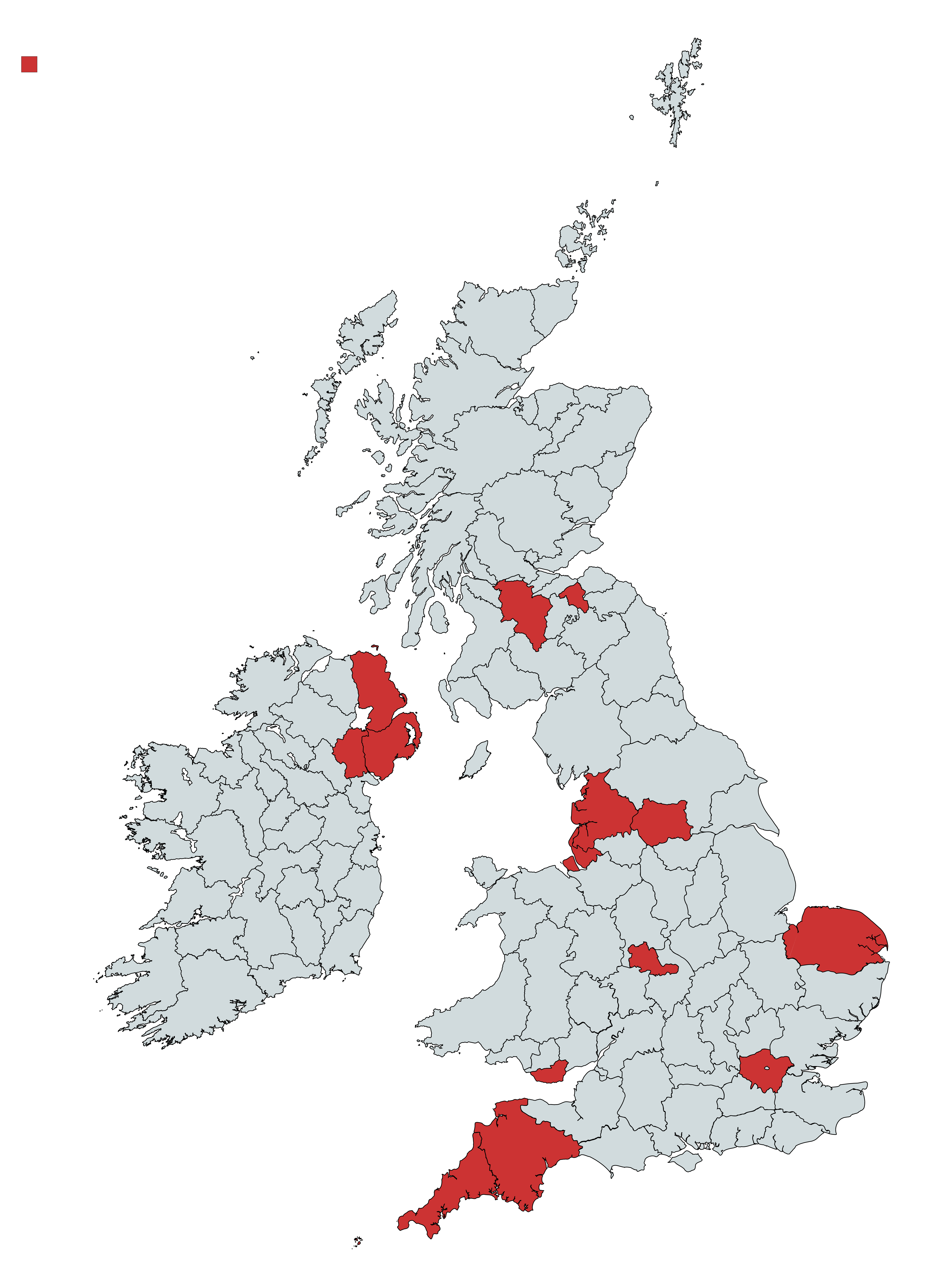}
        \caption{Geographical representation of locations used in training sets.}\label{fig:data_locations_on_map}
    \end{subfigure}
    \qquad
    \begin{subfigure}[b]{0.45\linewidth}
        \includegraphics[width=\linewidth, height=2.6cm]{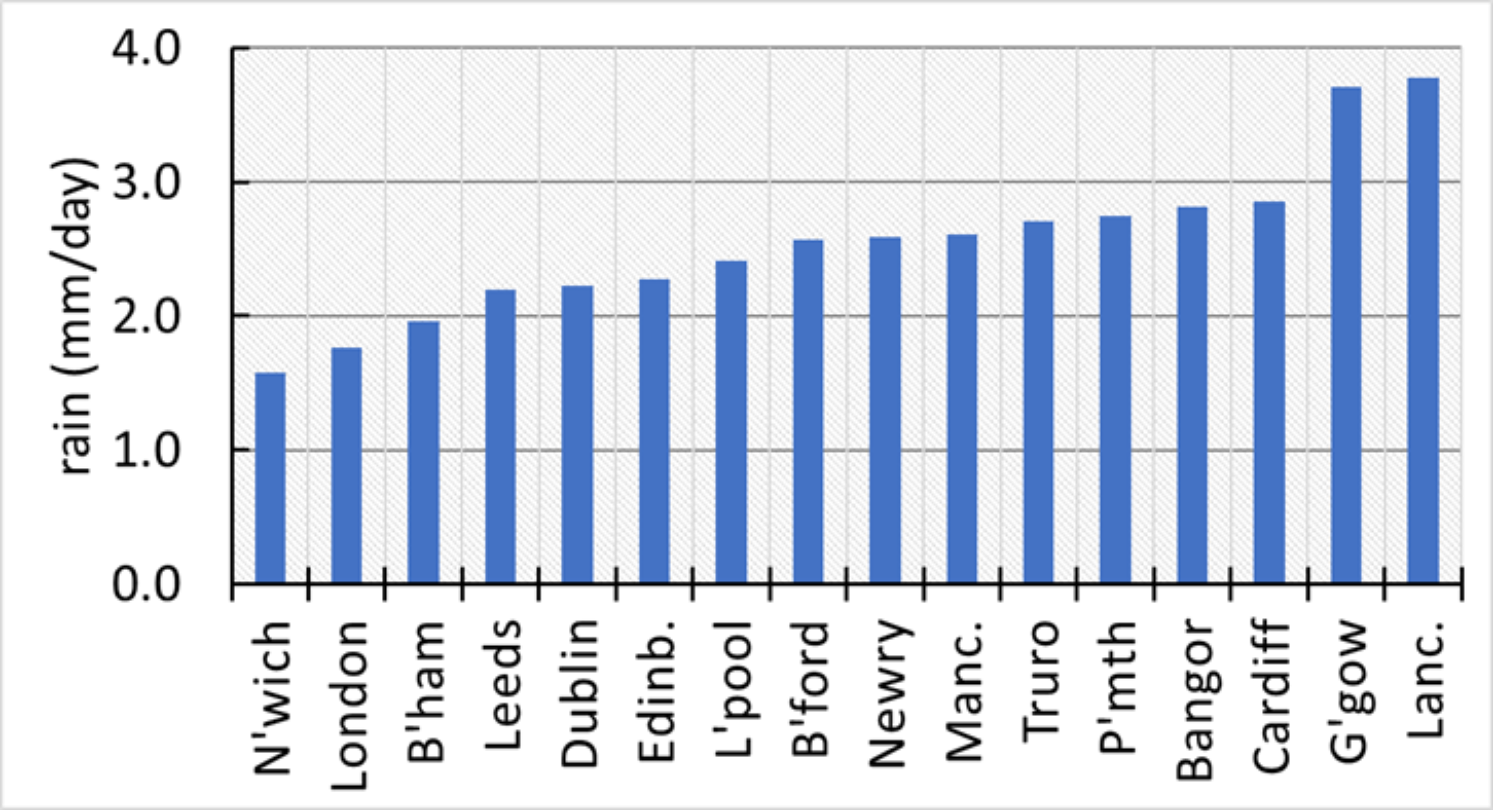}
        \caption{Average Daily Rainfall.}\label{subfig:AverageDailyRainfall}
    \vspace{2ex}
        \includegraphics[width=\linewidth, height=2.6cm]{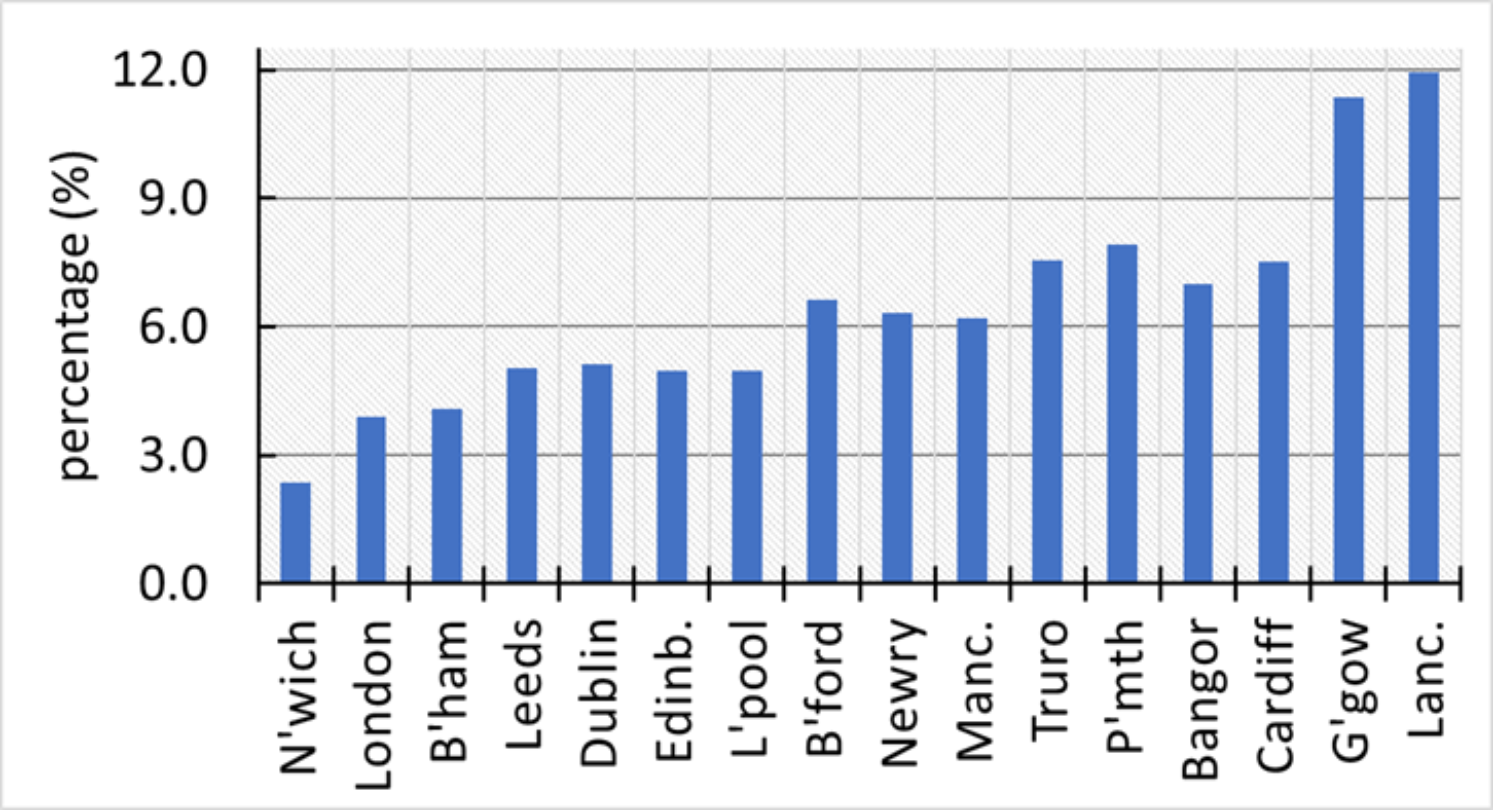}
        \caption{Percentage of Days with Rainfall larger than 10 mm/day. }\label{subfig:PercentageR10days}
   \vspace{2ex}
        \includegraphics[width=\linewidth, height=2.6cm]{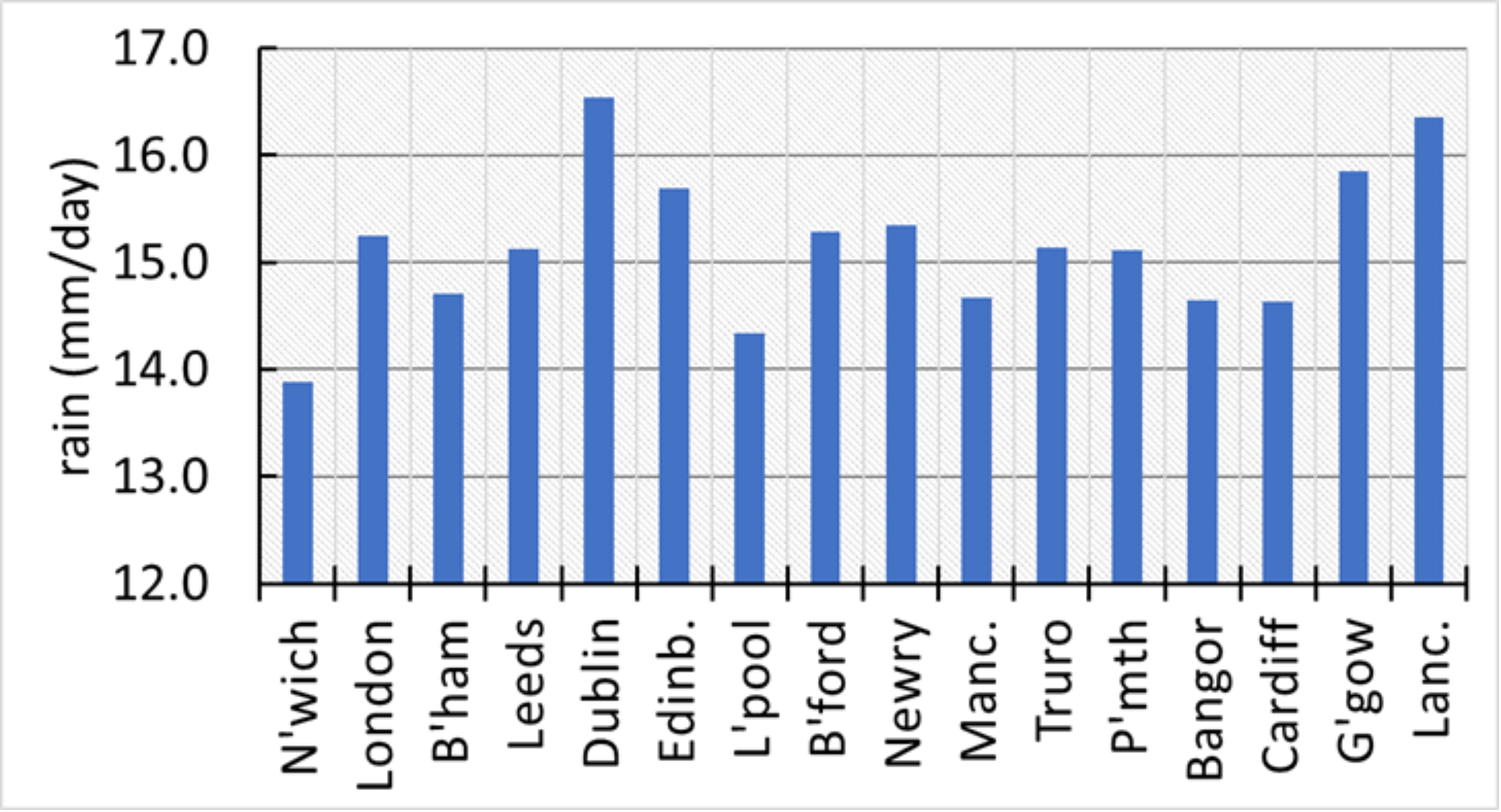}
        \caption{Average Daily Rainfall for R10 events.}\label{subfig:AverageR10rainfall}
  \end{subfigure}
  
\caption{\textbf{Precipitation profiles: } Here we present three statistics for each of the regions used in the training set for the experiment in Section~\ref{sec:EvaluationAgainstBaselines} where we evaluate our approaches against Baselines. }
\label{fig:precipitation_profiles}
\end{figure}


For training, we use the $16\times 16$ stencils surrounding sixteen locations to form our training and validation sets, namely, Cardiff, London, Glasgow (G'gow), Birmingham (B'ham), Lancaster (Lanc.), Manchester (Manc.), Liverpool (L'pool), Bradford (B'ford), Edinburgh (Edin), Leeds, Dublin, Truro, Newry, Norwich, Plymouth (P'mth) and Bangor. 

These locations were chosen as they are important population centres that sample a wide breadth of locations across the UK. Further, collectively these locations posses varied meteorological profiles, depicted in Figure~\ref{fig:precipitation_profiles}. For example, percentage of days with rainfall >10mm (R10) ranges from 2.4\% to 11.9\% and average rainfall conditional on an R10 event is ranging from 13.8 mm to 16.5 mm. 

During testing, we either test on the whole UK, region by region, or test on a the region around a single location such as a city. 

\section{Models}
\label{sec:models}

In the following we initially introduce our deep learning approach, TRU-NET as well as the Conditional Continuous loss used to train it. After this, we introduce the Hierarchical Convolutional Gated Recurrent Unit model.

\subsection{Temporal Recurrent U-NET}
\label{sec:trunet}

\begin{figure*}[htbp]
\centering
    \includegraphics[width=\textwidth]{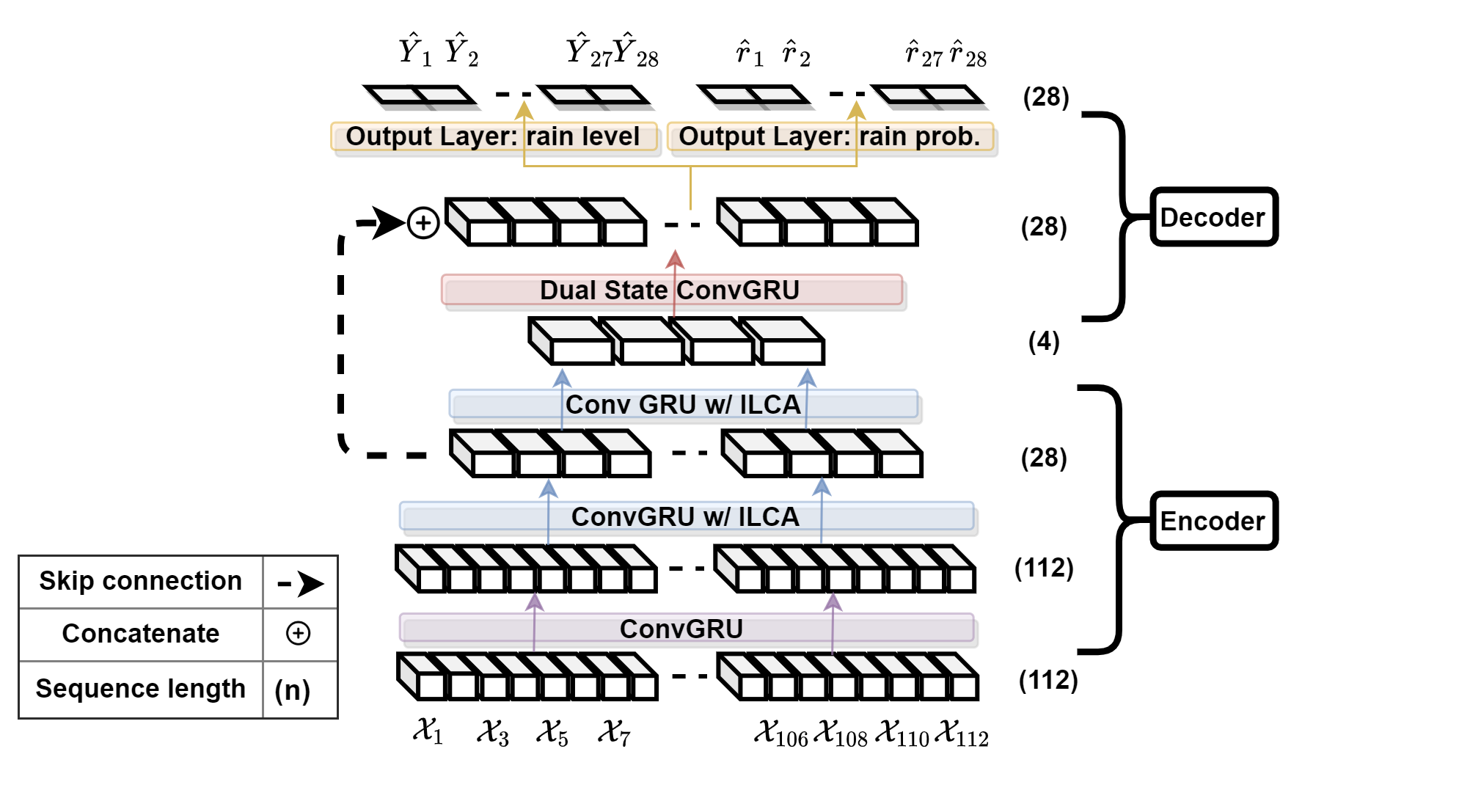}
  \caption{\textbf{TRU-NET architecture: }. This figure depicts the Conditional Continuous variant, which outputs predictions for rain level, $\hat{Y}_t$, and rain probability, $\hat{r}_t$, for 28 consecutive days. Each input tensor $\mathcal{X}_t$ is the concatenation of four time sequential $\mathcal{X}_{tj}$ matrices, presented in Figure~\ref{fig:model_input_output}. The Sequence Length of 3D tensors between layers contracts/expands through the encoder/decoder. This relates to an increasing/decreasing of the temporal scales modelled. The horizontal direction from left to right indicates time.}
\label{fig:TRUNETstructure}
\end{figure*}

Our TRU-NET model, visualised in Figure~\ref{fig:TRUNETstructure}, maps the 6-hourly low resolution model fields, $ \mathcal{X}_t \in \mathbb{R}^{4 {\times} 20 {\times} 21 {\times} 6}$, to a representation capturing variability on 6-hourly, daily and weekly time scales, and then uses these representations to output a prediction, $ \hat{Y}_t \in \mathbb{R}^{100 {\times} 140}$, for daily total rainfall.

As a first step within TRU-NET, we map the input data of the coarse grid onto the fine grid using bi-linear upsampling\footnote{Please note that an increase in resolution is called \emph{upsampling} in machine learning but \emph{down-scaling} in meteorology literature.}.

The encoder contains a stack of 3 bi-directional ConvGRU layers. Within the encoder, these 3 layers map the input into coarser spatial/temporal scales, from six-hourly/8.5km, to daily/34km, and to weekly/136km. To achieve this reduction in the temporal scales modelled by contiguous encoder layers, we propose a novel Fused Temporal Cross Attention mechanims (FTCA) as shown in Figure~\ref{fig:ConvGRU_w_FTCA}. These scales are aligned to the timescales associated with extreme rainfall events in the UK \citep{burton_2011}.

The decoder maps the latent representation captured at the weekly scale back to the daily scale before feeding it to an output layer for daily rain prediction.

Due to memory constraints we do not input the full (28$\cdot$4${\times}$100${\times}$140${\times}$6) dimensional model fields at once. In space, we extract stencils of $16\times 16$ grid-points for the input to predict precipitation over the stencil of $4\times 4$ grid-points in the centre of input stencil. TRU-Net processes 28 days worth of information at a time, generating an output of total daily precipitation for all of the 28 days for each application of TRU-NET:
\begin{equation}
    \hat{Y}_{t}, \ldots, \hat{Y}_{t-J}=f( \mathcal{X}_{t}, \mathcal{X}_{t-1}, \ldots,\mathcal{X}_{t-J} )
    \label{eqn: THST approach to Precip Forecasting}
\end{equation}
with $J=28$. This will naturally generate a lack of information on the past for the first timesteps ($J=1,2,3...$) and a lack of information on the future for the last timesteps ($J=...26,27,28$). However, this could be avoided by a stream of input data that only makes predictions for the time-steps in the centre of the time-series in future studies. 

In the following, we describe each of the main components of TRU-NET in more detail.

\subsubsection{Encoder}
\label{sec:encoder}

The encoder of our TRU-NET model, as shown in Figure~\ref{fig:TRUNETstructure}, has $L=3$ ConvGRU layers, where the $l$-th layer decreases the sequence length by a factor $m_l$: $ 1 \rightarrow 4 \rightarrow 7 $. This results in the number of units in each ConvGRU based layer decreasing in the manner: $112 \rightarrow 28 \rightarrow 4$, corresponding to six-hourly, daily and weekly temporal resolutions. 

The conventional ConvGRU is a recurrent neural network designed to model spatial-temporal information. In a conventional ConvGRU Layer, each unit $i$ shares its trainable weight matrices $\{W_k, U_k, b_k: k\in[z,r,\tilde{A}]\}$ with other units in the layer, and collectively they are described as having \textit{tied weights}. Each unit $i$ takes two inputs, namely the previous state $A_{i-1}$ and the input in the current time step $\widehat{\mathcal{B}_{i}}^{(h_b, w_b, c_b)}$, and outputs a state $A_{i}^{(h_a, w_a, c_a)}$, as detailed below. Here, $z_i$ is the update gate, $r_i$ is the reset gate, $\tilde{A}$ is the cell state, $\bullet$ and * denote the Hadamard product and convolution, respectively.
\begin{align}
\begin{split}
z_{i} &= \sigma\left(\widehat{\mathcal{B}_{i}} \ast W_{z} + A_{i-1} \ast U_{z} + b_{z}\right) \\
\tilde{A}_{i} &= \tanh \left(\widehat{\mathcal{B}_{i}} \ast W_{\tilde{A}} + r_{i} \bullet A_{i-1} \ast  U_{\tilde{ A}} + b_{\tilde{A}}\right) \\
r_{i} &= \sigma\left(\widehat{\mathcal{B}_{i}} \ast W_{r} + A_{i-1} \ast U_{r} + b_{r}\right) \\     
A_{i} &= z_i \bullet A_{i-1} + (1-z_i) \bullet \tilde{A}_{i}
\label{eqn:ConvGRU_equation}
\end{split}
\end{align}

When mapping an input from one time scale to another, e.g. generating the daily time scale tensor for day $t$ from a sequence of 4 corresponding six-hourly time scale tensors, a simple approach is to average the 4 six-hourly tensors. However, such a simple aggregation strategy ignores the influence of the daily time scale tensor from the previous day $t-1$. We instead propose \emph{Fused Temporal Cross Attention} (FTCA), as a better aggregation strategy based on the cross attention mechanism.

In the final two ConvGRU layers of the encoder, 
FTCA is fused into the ConvGRU in order to aggregate the inputs from the previous layer to generate a representation for the current layer. 
The ConvGRU with FTCA is illustrated in Figure~\ref{fig:ConvGRU_w_FTCA} and explained in the following subsection.

\subsubsection{Convolutional Gated Recurrent Unit with Fused Temporal Cross Attention (ConvGRU w/ FTCA) }
\label{sec:ConvGRUFTCA}

\begin{figure*}[htbp]
    \begin{minipage}{\textwidth}
    \centering
    \end{minipage}
    \vfill
    \begin{subfigure}[!t]{0.60\textwidth}
        \centering
        \includegraphics[trim={0 0 0 73},clip,width=\textwidth]{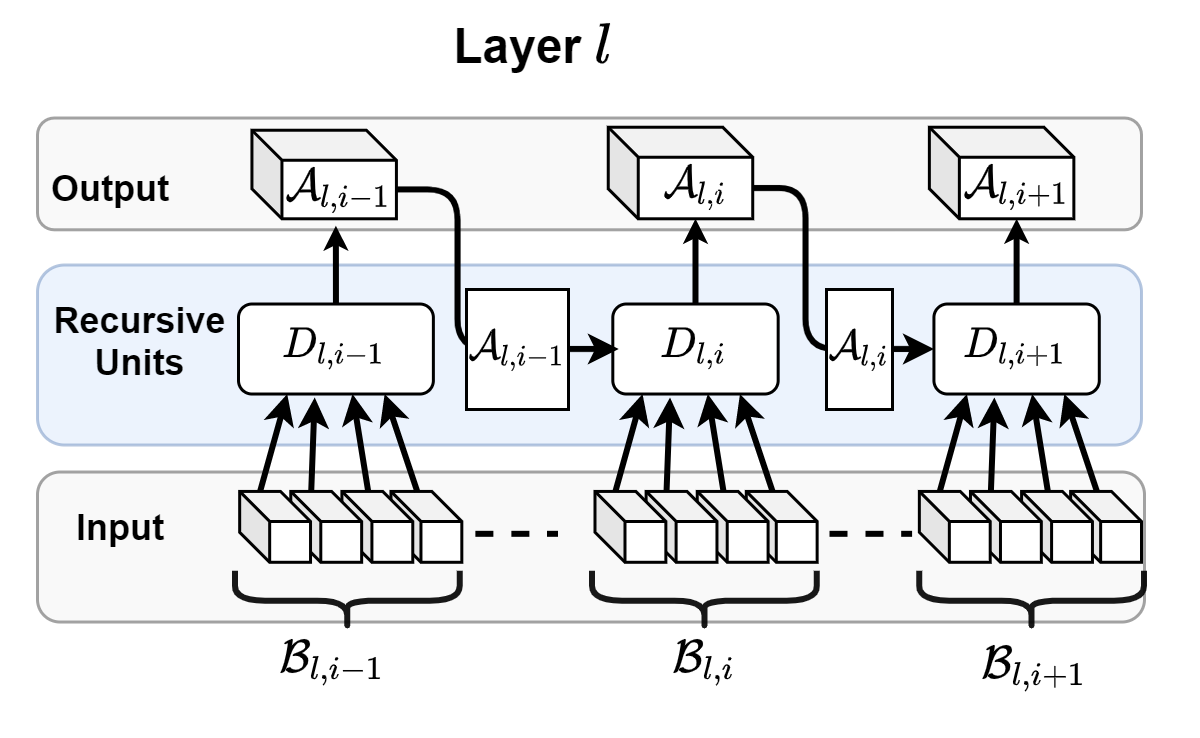}
        \caption{Layer $l$}
    \end{subfigure}
    \hfill
    \begin{subfigure}[!t]{0.40\textwidth}
        \centering
        \includegraphics[trim={0 30 0 72},clip,width=\textwidth]{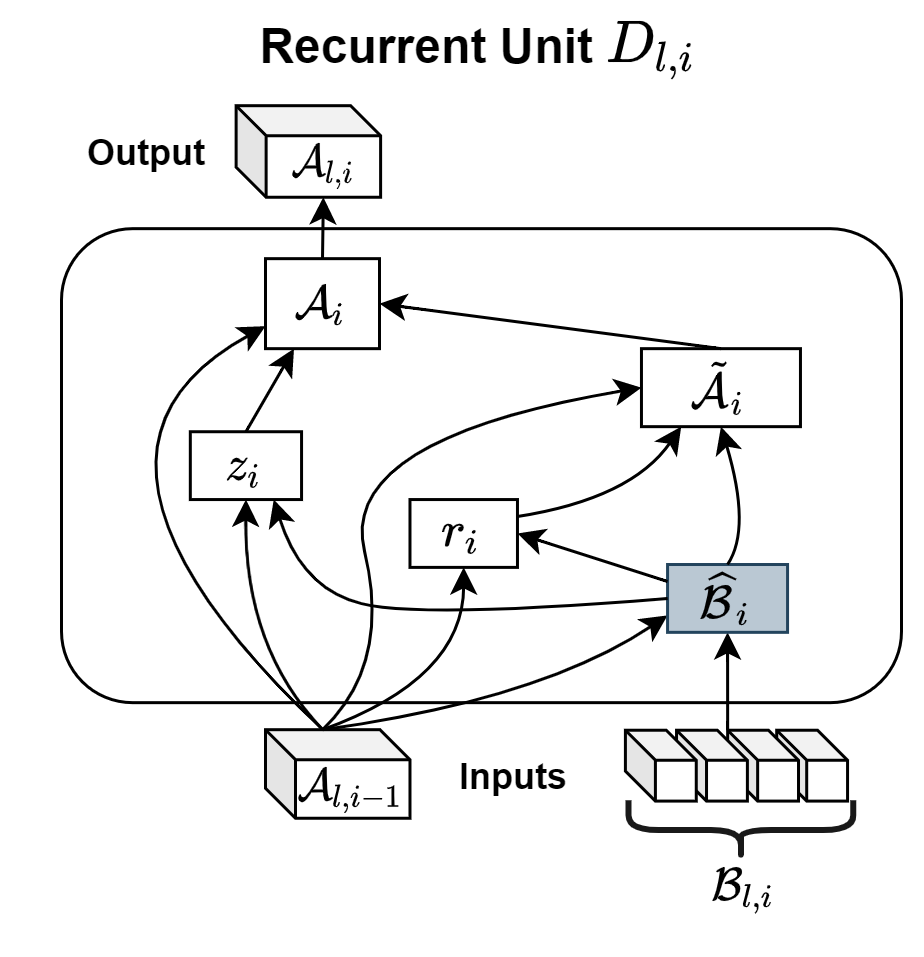}
        \caption{Recurrent Unit $D_{l,i}$}
    \end{subfigure}
      \caption{\textbf{ConvGRU with Fused Temporal Cross Attention (FTCA): } (a) illustrates a ConvGRU with FTCA layer and (b) illustrates an individual unit within the layer. The grey box in (b) shows our adaptation of the generic ConvGRU through the addition of an FTCA operation (grey box) that outputs $\widehat{\mathcal{B}}_i$. }
    \label{fig:ConvGRU_w_FTCA}
\end{figure*}

In the conventional ConvGRU, the $i^{th}$ unit of the $l^{th}$ layer, denoted as $D_{l,i}$, takes two inputs, the previous state $A_{i-1}$ and the input in the current time step $\widehat{\mathcal{B}_{i}}$. In our setup here, however, we stack ConvGRU layers with different temporal scales. As such, the input in the current time step to $D_{l,i}$ is no longer a single tensor, but instead, an ordered sequence of tensors, $\mathcal{B}_i \equiv B^{(h_b, w_b, c_b)}_{1:T_b}$, as shown in Figure \ref{fig:ConvGRU_w_FTCA}(a), where the input $\mathcal{B}_{l,i}$ consists of $T_b$ time-aligned outputs from the $(l{-}1)$-th ConvGRU layer, i.e., $\mathcal{B}_{l,i} \equiv \lbrace A_{l-1,j: j=1, \ldots, T_b} \rbrace$. For example, if the $l^{th}$ layer has the daily time resolution, then the $(l-1)^{th}$ layer would have the six-hourly time resolution, $T_b=4$ and $\mathcal{B}_{l,i} \equiv \lbrace A_{l-1,1}, A_{l-1,2}, A_{l-1,3}, A_{l-1,4} \rbrace$. 

Given $\mathcal{B}_i \equiv B^{(h_b, w_b, c_b)}_{1:T_b}$, we propose a Fused Temporal Cross Attention (FTCA) mechanism to calculated a weighted average $\widehat{\mathcal{B}_i}$. Here, we use $\mathcal{A}_{i-1}$ to derive a query tensor and $\mathcal{B}_i$ to derive both a key tensor and a value tensor. The query tensor is compared with the key tensor to generate weights which are used to aggregate various elements in the value tensor to obtain the final aggregated representation of $\widehat{\mathcal{B}_{i}}$.

Afterwards, the ConvGRU operations in Equation~\ref{eqn:ConvGRU_equation} are resumed. The FTCA related operations for unit $i$ have been decomposed into the following three steps:

\begin{itemize}[leftmargin=*]
    \item \textit{Downscaling representations: }On $A_i$ and $\mathcal{B}_i$, we first perform a 3D average pooling\footnote{The use of 3D average pooling is motivated by the high spatial correlation within a given feature map due to the spatially correlated nature of weather and to reduce the computational expense of the matrix multiplication.} (3DAP) with a pool size of ${M\times} M{\times} 1$ and transform them to matrices $\mathcal{A}_i^{PF}$ and $\mathcal{B}_i^{PF}$ of dimensions $(1, d_a)$ and $(T_b,d_b)$ respectively, via matrix-reshaping, where $d_a = h_a {\times} w_a {\times} c_a {\times} M^{-2}$ and $d_b = h_b {\times} w_b {\times} c_b {\times} M^{-2}$. 
\begin{align}
\begin{split}
    \mathcal{A}_i^{PF} &= \operatorname{Reshape} \big( \operatorname{3DAP} ( \mathcal{A}_i ) \big)^{(\frac{h_a}{M},         \frac{w_a}{M}, c_a) \rightarrow (1,d_a) } \\
    \mathcal{B}_i^{PF} &= \operatorname{Reshape} \big( \operatorname{3DAP} ( \mathcal{B}_i ) \big)^{(T_b,\frac{h_b}{M}, \frac{w_b}{M}, c_b) \rightarrow (T_b,d_b) } 
\label{eqn:CrossAttnMechanims_1}
\end{split}
\end{align}
\item \textit{Similarity calculation using relative attention score (RAS): }We transform $\mathcal{A}_i^{PF}$ and $\mathcal{B}_i^{PF}$ to $Q = \mathcal{A}_i^{PF} \circ W_Q$ and $K^{(T_b,d_o)} = \mathcal{B}_i^{PF} \circ W_K$ through matrix multiplication, $\circ $, with two trainable weight matrices, $W_Q^{(d_a,d_o)}$ and $W_K^{(d_b,d_o)}$. We then compute a matrix of weights $\mathcal{S}^{(1,T_b)}$, corresponding to the $T_b$ vectors in $K$, as follows:
\begin{equation}
    \mathcal{S}=\operatorname{softmax}\left(\frac{Q\circ(K + a^K)^T}{\sqrt{d_{o}}}\right) \label{eqn:similarity}
\end{equation}

Note here we use the relative attention score (RAS) function \citep{shaw2018_relattn} to compute the similarity in Equation~\ref{eqn:similarity}. Generally to calculate the similarity scores between $Q$ and each vector $K_b$, the inner product function is used \citep{Attention_is_all_you_need}. RAS extends this inner product scoring function by considering the relative position of each vector $K_b$ to one another. In our case, this position relates to the temporal position of $K_b$ relative to other members of $K$. To facilitate this, we also learn vectors $a_{b}^K$ which encode the relative position of each $K_b$.

\item \textit{Informative representation: }Finally the new informative representation $\widehat{\mathcal{B}}$ is learnt using two trainable convolution weight matrices with $c_f$ filters, $W_{V_1}^{( c_f , 4, 4, c_b)}$ and $W_{V_2}^{( c_f , 3, 3,c_b )}$ and a set of trainable vectors $a_{i}^V\in a^V$, encoding the relative position of each vector $V_i\in V$ as following: 
\begin{eqnarray}
V = \mathcal{B} \ast W_{V_1} \quad \quad   \widehat{\mathcal{B}} = (\mathcal{S} \circ (V + a^{v}) ) \ast W_{V_2}
\end{eqnarray}
\end{itemize}

We also use Multi-Head Attention (MHA) which allows the attention mechanism to encode multiple patterns of information by using $H$ heads, $\lbrace W_{Q},W_K,W_{V_1} \rbrace$, and performing $H$ parallel cross-attention calculations. The different values of $\lbrace W_{Q},W_K,W_{V_1} \rbrace$ across the heads capture different pattern/relationship in data, whereas simply using one head will lead to less diverse or informative patterns captured.  
\begin{align}
\begin{aligned}
    Q^{h} &= \mathcal{A}_{PF} \circ W^h_Q & Q &= \operatorname{Concat}(Q^h_{1:H}) \\
    K^{h} &=\mathcal{B}_{PF} \circ W^h_K &  K &= \operatorname{Concat}(K^h_{1:H}) \\
    V^{h} &= \mathcal{B} \ast W^h_{V_1} & V_i &= \operatorname{Concat}(V^h_{1:H}) \\
    \label{eq:Multi-head attention}
\end{aligned}
\end{align}

\subsubsection{Decoder}
\label{sec:decoder}
The decoder is composed of one Dual State ConvGRU (dsConvGRU) layer and an output layer which outputs predictions $\hat{Y}_{t:t+28}$ for the rain level for 28 consecutive days. If the conditional-continuous framework is in use, a second output layer outputs the corresponding predictions for the probability of rainfall $\hat{r}_{t:t+28}$ as illustrated in Figure~\ref{fig:TRUNETstructure}.

\paragraph{dsConvGRU: }
As illustrated in Figure~\ref{fig:TRUNETstructure}, the inputs to the dsConvGRU layer comes from the $2$nd and the $3$rd Encoder layers, 
while the output of the dsConvGRU layer is a sequence of 28 tensors which form a latent representation for the 28 days of the target observed daily precipitation $Y_{t:t+28}$.


As the dsConvGRU layer contains 28 units, we must expand the $3$rd Encoder layer's output from sequence length 4 to sequence length 28. To do this, we repeat every element in the sequence of length 4, 7 times, as in \citep{TreeStructureLSTM}. As such, each unit in the dsConvGRU layer receives an input from the temporally aligned unit in the $3$rd Encoder layer.

Extending Equations~\ref{eqn:ConvGRU_equation}, the dsConvGRU augments the conventional ConvGRU by replacing the input $\widehat{\mathcal{B}}_i$ with two separate inputs $\widehat{\mathcal{B}}_{i,(1)}$ and $\widehat{\mathcal{B}}_{i,(2)}$, each possessing the same dimensions as $\widehat{\mathcal{B}}_i$. Further, the $i$-th unit of the dsConvGRU layer takes three inputs, $A_{i-1}, \widehat{\mathcal{B}}_{i,(1)}$ and $\widehat{\mathcal{B}}_{i,(2)}$, and outputs a state $A_i$. 

Finally, referring to Equations~\ref{eqn:ConvGRU_equation}, we calculate two sets of the values, 
\newline $\{ z_{i,(j)}$,$r_{i,(j)}$,$\tilde{A}_{i,(j)}$,$A_{i,(j)}\}_{j \in [1,2]}$, corresponding to the use of $\widehat{\mathcal{B}}_{(1)}$ or $\widehat{\mathcal{B}}_{(2)}$ in place of $\widehat{\mathcal{B}}$. Finally, $A_i$ is calculated as the average of $A_{i,(1)}$ and $A_{i,(2)}$. 

\paragraph{Output Layer: } 
As we need to output two sequences of values, rainfall probabilities $\hat{r}_{t:t+28}$ and rainfall values $\hat{Y}_{t:t+28}$, for the conditional-continuous framework which will be discussed in Section \ref{sec:cc}, our model contains a separate output layer stacked over the dual-state ConvGRU layer for each output. 
Each output layer contains two 2D convolution layers, with 32 and 1 filters respectively and a kernel shape of (3,3).

\subsubsection{Conditional Continuous (CC) Augmentation}
\label{sec:cc}
To reflect the zero-skewed nature of rainfall data, due to many days without rainfall, a probabilistic conditional continuous (CC) distribution \citep{Husak2007_GammadistrtoreprRain,Stern1985_AModelFittingAnalysis} is often used to model precipitation. These distributions can be interpreted as the composition of a discrete component and a continuous distribution to jointly model the occurrence and intensity of rainfall:

\begin{align}
\delta(Y_t)&\approx\left\{\begin{array}{ll}
1, & Y_t=0 \\
0, & Y_t \neq 0
\end{array}\right.\\
p\left(Y_t;\gamma\right) &=
(1-r_t)\delta(Y_t) + r_t \cdot g(Y_t) \cdot (1-\delta(Y_t))
\label{eqn:rain_distr}
\end{align}

\noindent where $\delta$ is the Dirac function such that $\int_{-\infty}^{\infty} \delta(x) d x=1$, $r_t$ is the probability of rain at $t$-th day and $g(\cdot)$ is a Gaussian distribution with unit variance and predicted rainfall $\hat{Y}_t$ as mean. Therefore $(1-r_t)\delta(y_t)$ models the no rain events, while $r_t \cdot g(Y_t) \cdot (1-\delta(Y_t))$ handles the rain events. 



\citep{Vandal_2018} used a probabilistic CC loss for precipitation upscaling by using a Bayesian Neural Network (BNN) and performing multiple stochastic forward passes to generate a sample of predictions for each input. From this sample, estimates for the first and second moments are used to calculate the predictive loss and perform one back propagation step.
In contrast our model avoids the computational expense of multiple forward passes by outputting a prediction, $\hat{r}_t$, for the probability of rain occurring as well as a prediction, $\hat{Y}_t$, for the level of rainfall conditional on day $t$ being a rainy day.

To facilitate the requirement of two outputs, $\hat{Y}_t$ and $\hat{r}_t$, we augment the decoder to contain a second identical output layer. In this case, the TRU-NET model has a branch like structure, with $\hat{r}_t$ and $\hat{Y}_t$ the respective outputs of each of these branches.

During training, we sample one set of $[\hat{Y}_t,\hat{r_t}]$ per prediction and use the following loss function. This can be observed as a combination of the binary cross entropy on predictions for whether or not it rained (the first term) and a squared error term  on the predicted unconditional rainfall intensity (the second term).

\begin{equation}
\begin{split}
\mbox{L}(Y_t,[\hat{Y}_t,\hat{r}_t]) = \frac{1}{T}\Bigg[\sum_{t=1}^T \bigg(
\mathbbm{1}_{\mathbf{y}_{t}>0} \cdot \log({\hat{r}}_{t}) +  \left(\mathbbm{1}_{\mathbf{y}_{t}=0}\right) \cdot\log\left(1-\hat{r}_{t}\right)\bigg) \\
- \sum_{t=1}^{T} \left\|Y_t-\hat{Y}_{t} \right\|^{2}\Bigg] \label{eqn:loss_function}
\end{split}
\end{equation}

\subsubsection{Monte Carlo Model Averaging (MCMA) }
\label{sec:MCMA}
When training with dropout, each of the $n$ weights in the neural network has a probability $p$ of being masked. As such, there are $2^n$ possible models, defined by the combination of weights that can be masked. When sampling predictions from the model, it is infeasible to sample from each of the $2^n$ variations. Instead, we can form a sample of predictions from a random selection of the $2^n$ possible models, and calculate the average of the sample. More formally, MCMA is the process of using dropout during training and testing. During training, dropout is performed with a fixed probability $p$ of masking weights. During testing we draw $n$ samples, from our model for each prediction. To do this we use $n$ different dropout masks on the model's weights. Each dropout mask uses the same masking probability, $p$, on the model's weight as was used during training.
We then calculate the mean of these samples to arrive at a \textit{model averaged} prediction. Experiments in \citep[\S 7.5]{Dropout_srivastava}, show this method is effective to sample from neural networks trained with dropout.

During inference, we use the MCMA 
framework to produce $i\in I$ samples $[\hat{r}^{i}_t, \hat{Y}^{i}_t]$ for each observed rainfall $Y_t$. For each observation, we calculate a final prediction $\hat{Y}_t$ for $Y_t$:
\begin{equation}
            \hat{Y}_t= \frac{1}{I} \sum_{i=1}^I  \mathbbm{1}_{\hat{r}^{i}_t>0.5} \cdot \hat{Y^{i}_t}
\end{equation}

\subsection{Hierarchical Convolutional GRU (HCGRU): } 
\label{sec:hcgru}

\begin{figure}[htbp]
\centering
    \includegraphics[trim={0 83 0 42},clip,width=0.7\linewidth]{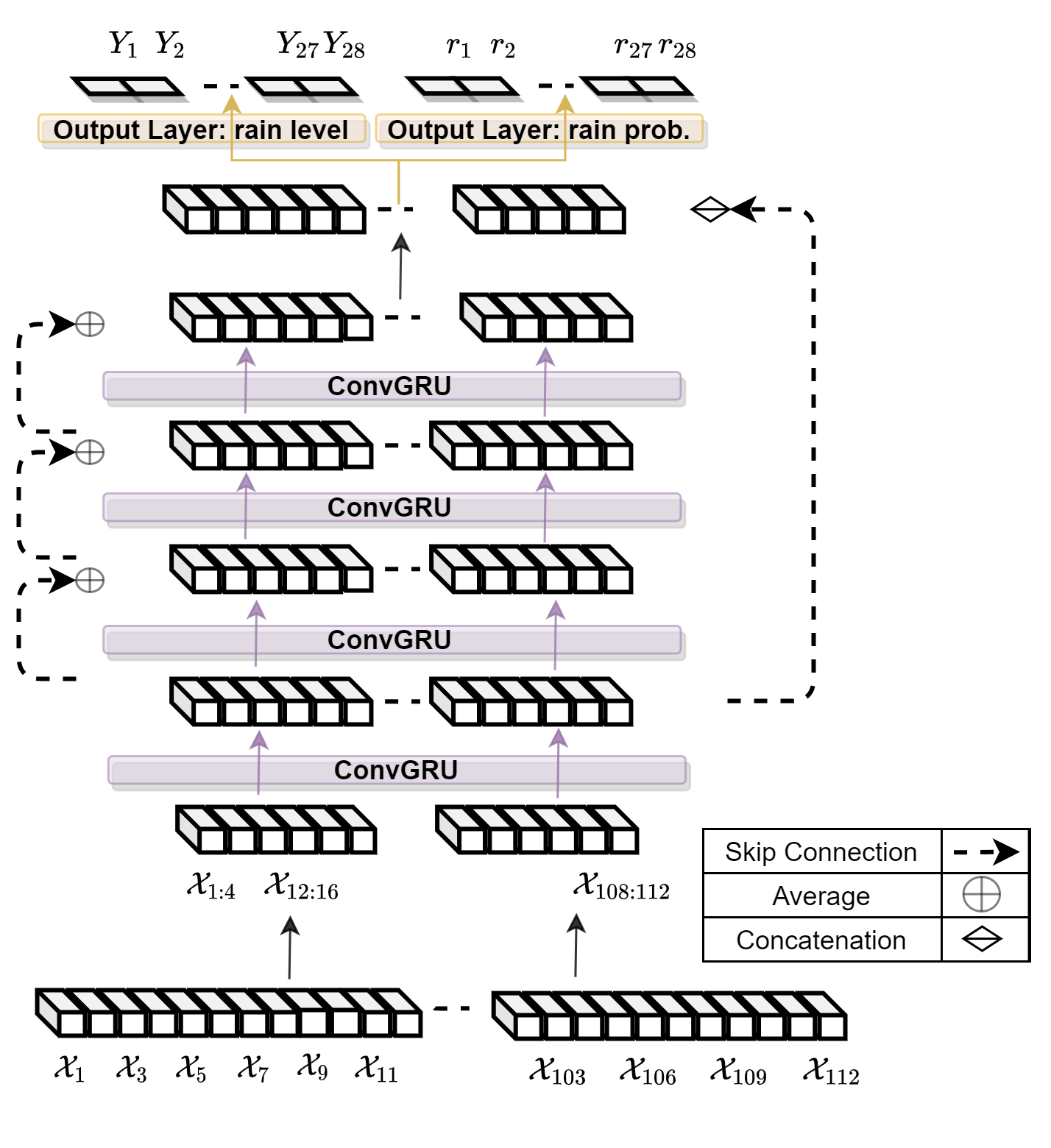}
    \caption{Illustration of the conditional-continuous variant of the HCGRU model used as a baseline in experiments.}
    \label{fig:HCGRU}
\end{figure}
The general structure of a stack of convolutional layers has been used successfully in precipitation nowcasting \citep{ConvLSTM_shi,shi2017deep} wherein it outperformed an Optical Flow algorithm \citep{Rover} produced by the Hong Kong Observatory. In the aforemention work, the task was sequence to point forecasting.
For this work, we adapt the structure to the task of sequence to sequence transduction and propose the structure illustrated in Figure~\ref{fig:HCGRU}. Our implementation contains 4 ConvGRU layers and an output layer, matching the number of layers in our TRU-NET model. Prior to the first layer, we reduce the input sequence from length 112 to 28, by concatenating blocks of 4 sequential elements. Each of the 4 ConvGRU layers contain 28 recurrent units, with each recurrent unit in each layer containing convolutional operations with 80 filters and a kernel shape of $(4,4)$. Skip connections exists over the final 3 ConvGRU layers and a final skip connection exists from the output of the first ConvGRU layer to the input of the output layer.
The output layer follows the same formulation as in TRU-NET, with two 2D Convolution layers. Similar to TRU-NET, we create two variants of HCGRU. One variant uses the Conditional Continuous loss and therefore has two output layers for probability of rain and conditional rain intensity. The other variant only outputs a prediction for rain intensity and as such only has one output layer.

\section{Experimental Setup}
\label{sec:training}

This section describes baseline models used for comparison, hardware setup.

\subsection{Baseline Models}
\label{sec:baselinemodels}

We compare TRU-NET with the following baselines:

\noindent \paragraph{Integrated Forecast System (IFS): } The IFS is a computationally expensive numerical weather prediction system which is solving the physical equations of atmospheric motion. IFS is used for operational weather predictions at the European Centre for Medium-Range Weather Forecasts (ECMWF). It is also used to generate the ERA5 reanalysis data which is used as input data for TRU-NET. While the input fields are a product of the data assimilation process of ERA5, there are also data for precipitation predictions available which are diagnosed from short-term forecast simulations with IFS which use ERA5 as initial conditions. There are two forecast simulations started each day at 6 am and 6 pm. We extract the precipitation fields for the first 12 hours of each simulation to reproduce daily precipitation - this is presently the optimal way to derive meaningful precipitation predictions from a dynamical model that is consistent with the large-scale fields in the ERA5 reanalysis data. The ERA5 and precipitation data is available on a grid with 31 km resolution. However, our target is to use model fields from climate models as input which are typically run at coarser resolution. We therefore map the ERA5 data onto the grid that is used in the HadGEM3 climate model \citep{Murphy2018}.

\noindent\paragraph{U-NET: } 
U-NET \citep{ronneberger2015unet} is a popular convolution based network architecture initially proposed for biomedical imaging segmentation, but recently used in image super-resolution and precipitation downscaling \citep{agrawal2019machine, trebing2020smaatunet}. Similar to TRU-NET and HCGRU, U-NET features an encoder and a decoder. However, unlike HCGRU and TRU-NET, U-NET does not feature any structure to capture temporal relationships. As U-NET is a simpler architecture, we have utilised a larger structure with 4 sub-layers in the Encoder and 4 sub-layers in the decoder. This ensures that the number of parameters present in U-NET is similar to that of TRU-NET and HCGRU, as shown in Table~\ref{tab:model_sizes}.

\begin{table}[htbp]
\centering
\begin{tabular}{@{}lllll@{}}
\toprule
                                                           & TRU-NET    & HCGRU     & U-NET      \\ \midrule
\begin{tabular}[c]{@{}l@{}}Parameter \\ Count\end{tabular} & 6,890,290 & 6,609,602 & 7,709,888 \\ \bottomrule
\end{tabular}
\caption{\textbf{Model Size: } Here we present the sizes of models used in our experiments. We choose similar sizes for the neural network architectures to ensure that any difference in performance is primarily due to modelling capability of the architecture as opposed to model size.}
\label{tab:model_sizes}
\end{table}

\subsection{Hyperparameter Settings}
\label{sec:hyperparameterSettings}

In this section we describe the hyper-parameter tuning process we used for the TRU-NET and HCGRU variants. We focus on the hyper-parameter tuning process for TRU-NET CC. An identical process was followed for the other models.

Initial training runs were used to guide our decision to tune the following hyper-parameters: input dropout, recurrent dropout, attention dropout, learning rate, $\beta_1$ and $\beta_2$ and clip norm. 
Input dropout and recurrent dropout are hyper-parameters used by the convolutional-recurrent based structures in the HCGRU and TRU-NET variants. Attention dropout is the hyper-parameter used by TRU-NET's attention based structures. The Rectified Adam optimizer's \citep{RectifiedAdam2020} hyper-parameters we chose to optimize were learning rate, $\beta_1$ and $\beta_2$. We also included clip norm for the gradients. 
Table~\ref{tab:hypertune_paramrange} show the values used for the hyper-parameter tune grid search. The dataset used for tuning includes the following  cities: Cardiff, London, Glasgow, Birmingham, Lancaster, Manchester, Liverpool, Bradford, Edinburgh, Leeds. The training set covered data from 1999 till 2008 and the test set contained data from 2009 till 2014.

\begin{table}[htbp]
    \begin{subtable}[h]{\textwidth}\centering\tabcolsep=0.125cm{
    
        \centering
        \begin{tabular}{rrrrrrr}
        \toprule
                   $\beta_2$ & learning rate &  input d.o. &  recurrent d.o. &  clip norm &  attention d.o. \\
            \midrule
                 0.900 &  1\text{e-}3 &      0.150 &       0.150 &     4.500 &   0.150 \\
               0.990 &  1\text{e-}4 &      0.350 &       0.350 &     5.500 &   0.350 \\
        \bottomrule
        \end{tabular}}
        \caption{TRU-NET CC}
        \label{tab:hypertune_paramrange_TRUNETCC}
    \end{subtable}
    \vfill
    \vspace*{3mm}
    \begin{subtable}[h]{\textwidth}\centering\tabcolsep=0.125cm{
        \centering
        \begin{tabular}{rrrrrr}
            \toprule
              $\beta_1$ &     $\beta_2$ &  learning rate & input dropout &  recurrent dropout \\
            \midrule
                 0.750 & 0.900 & $1\text{e-}3$ &      0.100 &       0.100 \\
                0.900 & 0.990 &   $1\text{e-}4$ &    0.225 &       0.225 \\
              &     &       &0.350 &       0.350 \\
            \bottomrule
        \end{tabular}}
        \caption{HCGRU CC}
        \label{tab:hypertune_paramrange_HCGRUCC}
    \end{subtable}
 \caption{\textbf{Hyper-parameter tuning values: } Here we present the values used for hyper-parameter grid search tuning. Sub-Table a) shows TRU-NET CC's values and Sub-Table b) shows HCGRU's values. The combination of these hyper-parameter options resulted in a set of 64 models for each of TRU-NET CC and HCGRU CC. d.o. is short-hand for dropout.}
\label{tab:hypertune_paramrange}
\end{table}

Our initial training runs indicated that TRU-NET CC's performance was significantly sensitive to clip norm. A clip norm below 6.5 and linear warmup were required to ensure stable training.

\begin{table}[htbp]
    
    \begin{subtable}[h]{\textwidth}\centering\tabcolsep=0.125cm{    
        \centering
        \begin{tabular}{rrrrrrrrr}
        \toprule
            $\beta_2$ & learning rate &  input d.o. &  recurrent d.o. &  clip norm &  attention d.o. &  R10 RMSE \\
            \midrule
             0.900 & 1\text{e-}4 &        0.350 &        0.150 &      4.500 &    0.350 &     6.410 \\
             0.990 & 1\text{e-}4 &        0.350 &        0.150 &      5.500 &    0.350 &     6.571 \\
             0.900 & 1\text{e-}4 &      0.150 &        0.150 &      5.500 &    0.350 &     6.744 \\
             0.990 & 1\text{e-}4 &       0.150 &        0.150 &      4.500 &    0.350 &      6.859 \\
             0.990 & 1\text{e-}3 &        0.150 &        0.350 &      5.500 &    0.350 &     6.894 \\
        \bottomrule
        \end{tabular}}
        \caption{TRUNET CC}
        \label{tab:R10RMSE_hypertuning_trunetcc}
    \end{subtable}    
    \vfill
    \vspace{3mm}
        \begin{subtable}[h]{\textwidth}\centering\tabcolsep=0.125cm{    
            \centering
            \begin{tabular}{rrrrrrrrr}
            \toprule
                 $\beta_1$ &    $\beta_2$ & learning rate &  input d.o &  recurrent d.o. &  R10 RMSE \\
                \midrule
                   0.900 & 0.900 & 1\text{e-}3 &       0.225 &        0.350 &     7.593 \\
                   0.900 & 0.990 & 1\text{e-}3 &       0.225 &        0.350 &     7.607 \\
                   0.750 & 0.990 & 1\text{e-}3 &       0.350 &        0.225 &     7.619 \\
                   0.750 & 0.990 & 1\text{e-}3 &       0.100 &        0.225 &     7.630 \\
                   0.900 & 0.990 & 1\text{e-}3 &       0.100 &        0.100 &     7.642 \\
            \bottomrule
            \end{tabular}}
            \caption{HCGRU CC}
            \label{tab:R10RMSE_hypertuning_hcgrucc}
        \end{subtable}
    \caption{\textbf{Grid search hypertuning results: } Here we present the five hyper-parameter settings that resulted in the best R10 RMSE scores for TRU-NET CC and HCGRU CC. Sub-Table a) shows the results for TRU-NET CC models trained for 40 epochs and Sub-Table b) shows the results for HCGRU CC models trained for 30 epochs. The training set consisted of data between 1999 to 2014 and the testing set consists of data between 2014 to 2019. d.o. is short-hand for dropout.}
    \label{tab:R10RMSE_hypertuning}
\end{table}

\begin{figure}[htbp]
\begin{subfigure}{1\linewidth}
  \centering
  \includegraphics[width=0.49\linewidth]{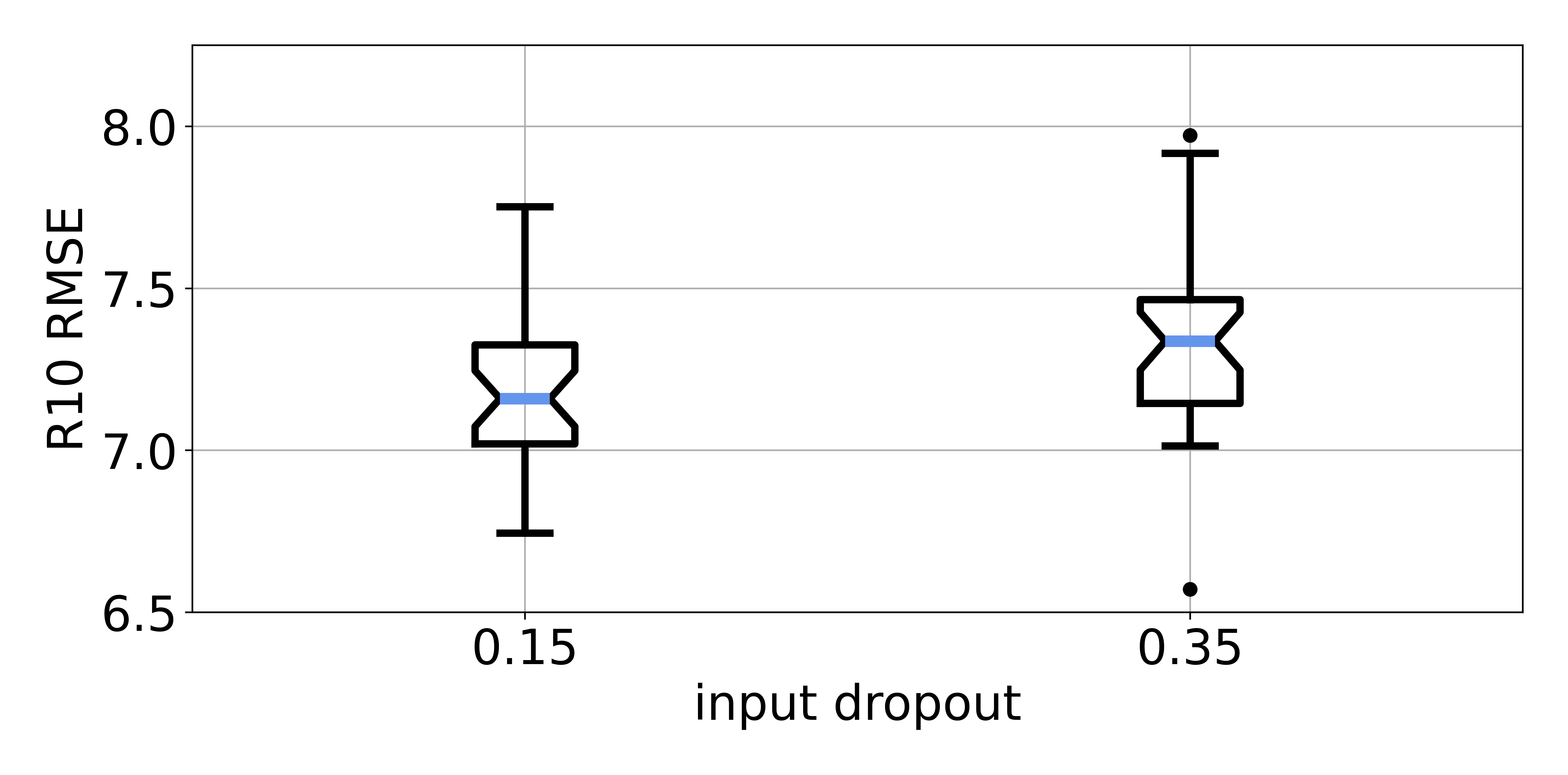}
    \caption{R10 RMSE to attention dropout}
\end{subfigure}
\vfill
\vspace*{2mm}
\begin{subfigure}{0.45\linewidth}
  \centering
  \includegraphics[width=1\linewidth]{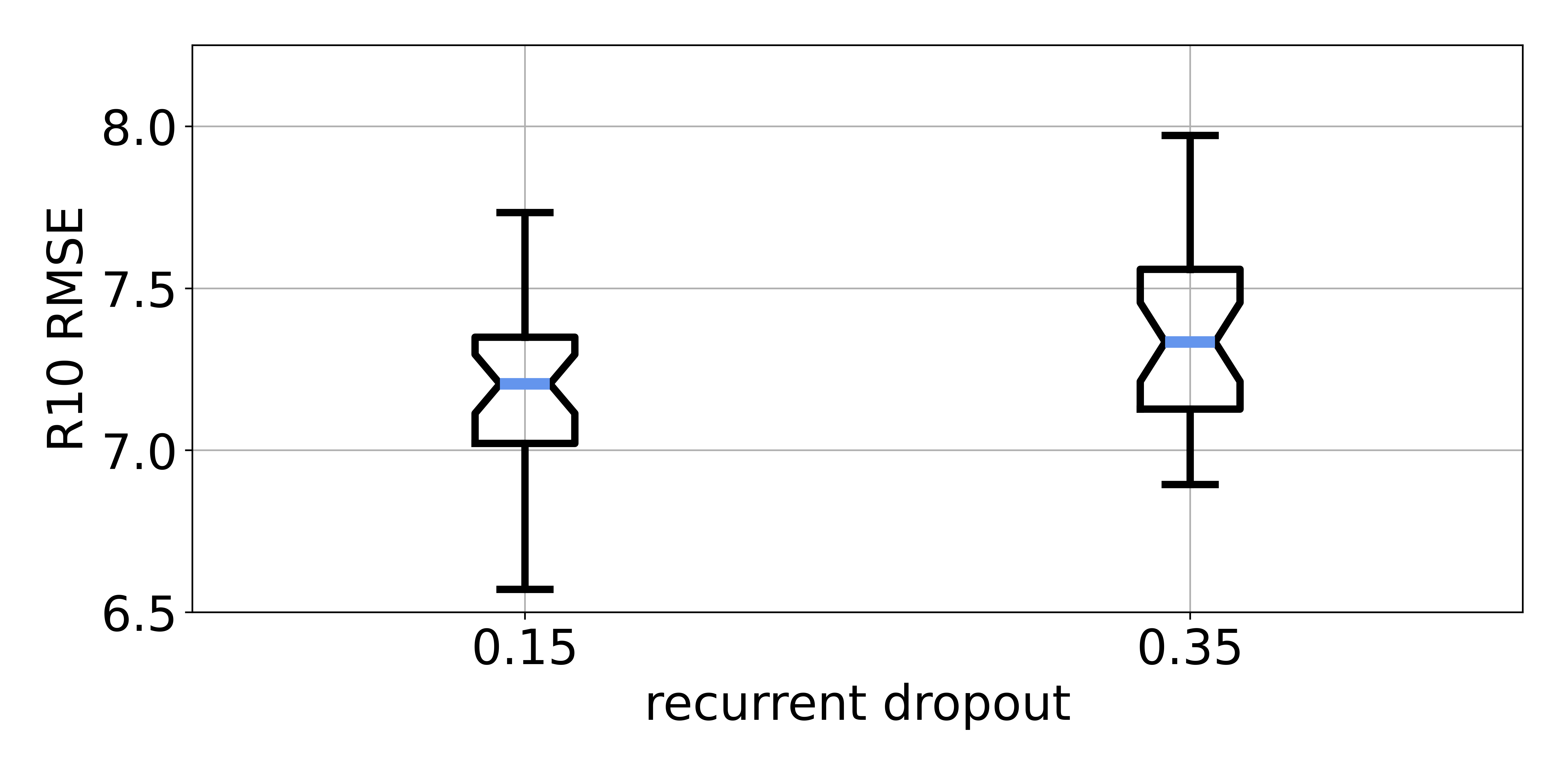}
    \caption{R10 RMSE to input dropout}
\end{subfigure}
\hfill
\begin{subfigure}{0.49\linewidth}
  \centering
  \includegraphics[width=1\linewidth]{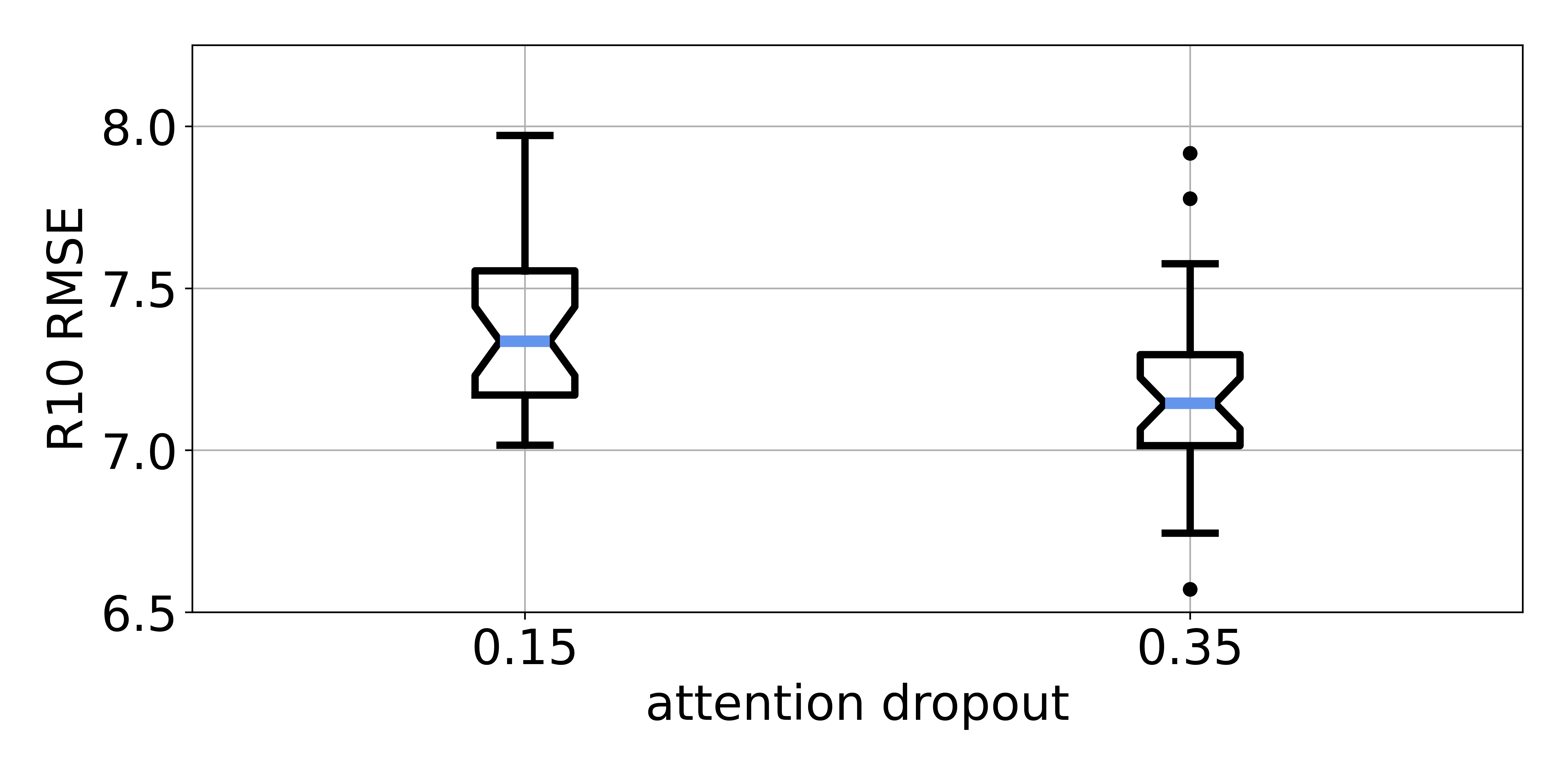}
    \caption{R10 RMSE to recurrent dropout}
\end{subfigure}
 \caption{\textbf{Hyperparameter tuning results: } Here we present the results of the 64 TRU-NET CC models that were trained. The Sub-Figures show the distribution of R10 RMSE values relative to the hyper-parameter values for the 64 TRU-NET models, judged by R10 RMSE. The units for R10 RMSE are mm/day. }
 \label{fig:TRUNET_CC_hypertuning_boxplots}
\end{figure}

Table~\ref{tab:R10RMSE_hypertuning} displays the hyper-parameter settings corresponding to the top 5 performing TRU-NET CC and HCGRU CC models, ranked by R10 RMSE.
For the TRU-NET CC models, the hyper-parameters with the most significant effect on R10 RMSE were input dropout, recurrent dropout and attention dropout. For these hyper-parameters Figure~\ref{fig:TRUNET_CC_hypertuning_boxplots} uses boxplots to illustrate the distribution of R10 RMSE scores conditional on hyper-parameter values. We see that values of 0.35, 0.15 and 0.15 for the attention dropout, input dropout and recurrent dropout optimize TRU-NET CC performance. Given the results presented in Table~\ref{tab:R10RMSE_hypertuning}, we settle on using the following hyper-parameter values for TRU-NET CC during experiments: learning rate=1\text{e-}4, $\beta_2$=0.90, clip norm=4.5, input dropout=0.15, recurrent dropout=0.15 and attention dropout=0.35. A similar procedure was followed for our HCGRU model to arrive at an optimal parameter range of learning rate=0.001, $\beta_1$=0.90, $\beta_2$=0.99, input dropout=0.225 and recurrent dropout=0.35.

\section{Experiments}
\label{sec:experiment}

\subsection{Comparison with Baselines}
\label{sec:EvaluationAgainstBaselines} 

\subsubsection{Seasonal breakdown for all of the UK}
\label{sec:seasonal_breakdown}
We use the following metrics to evaluate the performance of each model: Root Mean Squared Error (RMSE), RMSE for days of observed rainfall over 10/mm (R10 RMSE) and Mean Absolute Error (MAE). We present these metrics for each season, where the seasons have been defined as Spring (March, April, May), Summer (June, July, August), Autumn (September, October, November) and Winter (December, January, February). The training set spans the years 1979 till 2008, the validation set spans the years 2009 till 2013 and the test set spans the time period 2014 till August 2019. 
\begin{table}[h!]
\begin{subtable}{\linewidth}\centering{
        \begin{tabular}{lccc}
        \toprule
         Model name &  RMSE &  R10 RMSE &   MAE \\
        \midrule
        IFS        & 3.627 &     9.001 & 1.976 \\
        U-NET      &3.790  &9.677      &2.879  \\
        U-NET.CC      &3.580  &10.262      &1.846  \\
        HCGRU      & 3.268 &     8.792 & 1.762 \\
        HCGRU.CC   & 3.266 &     \textbf{8.671} & 1.739 \\
        TRU-NET      & 3.106 &     8.766 & 1.784 \\
        TRU-NET.CC   & \textbf{3.081} &   8.759 & \textbf{1.644} \\
        \bottomrule
    \end{tabular}}
    \caption{All Seasons}
\end{subtable}
\vspace*{3mm}
\begin{subtable}{.5\linewidth}\centering\tabcolsep=0.115cm{    
        \begin{tabular}{lccc}
        \toprule
        Model name &  RMSE &  R10 RMSE &   MAE \\
        \midrule
        IFS        & 3.950 &     9.114 & 2.233 \\
        U-NET        & 4.194  &9.929      & 3.021  \\
        U-NET.CC      & 4.144  & 10.651     & 2.252  \\
        HCGRU      & 3.740 &     \textbf{8.879} & 2.135 \\
        HCGRU.CC   & 3.731 &     8.894 & 2.039 \\
        TRU-NET      & 3.613 &     9.138 & 2.126 \\
        TRU-NET.CC   & \textbf{3.570} &     9.096 & \textbf{1.978} \\
        \bottomrule
    \end{tabular}}
    \caption{Winter}
\end{subtable}
\hfill
\begin{subtable}{.5\linewidth}\centering\tabcolsep=0.115cm{    
    \begin{tabular}{lccc}
        \toprule
        Model name &  RMSE &  R10 RMSE &   MAE \\
        \midrule
        IFS        & 3.135 &     8.455 & 1.692 \\
        U-NET      &3.355  &8.776      &2.719  \\
        U-NET.CC      &2.945  & 9.420     &1.483  \\        
        HCGRU      & 2.707 &     7.922 & 1.439 \\
        HCGRU.CC   & 2.710 &     7.832 & 1.419 \\
        TRU-NET      & 2.549 &     7.817 & 1.487 \\
        TRU-NET.CC   & \textbf{2.504} &     \textbf{7.777} & \textbf{1.328} \\
        \bottomrule
    \end{tabular}}
        \caption{Spring}
\end{subtable}
\vspace*{3mm}
\begin{subtable}{.5\linewidth}\centering\tabcolsep=0.115cm{
    \begin{tabular}{lccc}
        \toprule
        Model name &  RMSE &  R10 RMSE &   MAE \\
        \midrule
        IFS        & 3.663 &     9.021 & 2.018 \\
        U-NET      &3.646  &9.593      &2.850  \\
        U-NET.CC      &3.406  &9.940      &1.761  \\        
        HCGRU      & 3.210 &     9.056 & 1.718 \\
        HCGRU.CC   & 3.193 &     \textbf{8.701} & 1.695 \\
        TRU-NET      & 3.001 &    8.764 & 1.749 \\
        TRU-NET.CC   & \textbf{2.991} &     8.800 & \textbf{1.616} \\
        \bottomrule
    \end{tabular}}
    \caption{Summer}
\end{subtable}
\hfill
\begin{subtable}{.5\linewidth}\centering\tabcolsep=0.115cm{   
    \begin{tabular}{lccc}
        \toprule
        Model name&  RMSE &  R10 RMSE &   MAE \\
        \midrule
        IFS        & 3.765 &     9.222 & 1.987 \\
        U-NET      & 3.941 &10.011      &2.939  \\
        U-NET.CC      & 3.755  &10.532      &1.9311  \\        
        HCGRU      & 3.381 &     9.073 & 1.783 \\
        HCGRU.CC   & 3.398 &     8.923 & 1.773 \\
        TRU-NET      & \textbf{3.210} &     \textbf{8.892} & 1.798 \\
        TRU-NET.CC   & 3.215 &     8.926 & \textbf{1.680} \\
        \bottomrule
    \end{tabular}}
 \caption{Autumn}
\end{subtable}
    \caption{ \textbf{Comparing TRU-NET with Baselines: }. Here, we present the results for models tested on the whole UK between the dates of 2014 and August 2019. The TRU-NET CC (TRU-NET.CC) model achieves the best RMSE and MAE scores across all seasons. The units for all metrics are mm/day.}
    \label{tab:seasonal_breakdown}
\end{table}

In Table~\ref{tab:seasonal_breakdown}(a) we observe that the TRU-NET CC model generally outperforms alternative models in terms of RMSE and MAE. Further, our CC variants of TRU-NET and HCGRU achieve a better R10 RMSE than their non conditional continuous counterparts.

We notice that while TRU-NET CC achieves a lower R10 RMSE than HCGRU CC, HCGRU CC achieves a slighlty lower R10 RMSE than TRU-NET CC. However, the improvement (5.7\%) in the TRU-NET CC RMSE score relative to HCGRU CC RMSE score is far larger than the improvement (1.1\%) of the in the HCGRU CC R10 RMSE score relative to the TRU-NET CC RMSE score. As such we believe that a more exhaustive fine-tuning processing that tested more values for TRU-NET's dropout, in the FTCA structure, would lead to further improvements in the R10 RMSE for TRU-NET CC. Finally, we notice, that U-NET/U-NET CC achieves a significantly worse RMSE/R10 RMSE score than the other DL approaches. The poor relative performance of U-NET highlights the importance of using a a neural network structure that utilises the temporal information present in weather data.

\subsubsection{City-wise breakdown}
\label{sec:city_wise_breakdown}
In the previous sub-section, we presented seasonal performance metrics for each model tested on the whole country. Here we focus on the predictive errors on 5 specific cities across the range of precipitation profiles displayed in Figure \ref{fig:precipitation_profiles}. These cities chosen can be divided into two groups; those with lower rainfall characteristics (London, Birmingham and Manchester) and those with high rainfall characteristics (Cardiff and Glasgow). These locations have been chosen in order to discern whether the quality of predictions over a region is related to the region's precipitation profile. The following tables present the predictive scores of the TRU-NET CC model trained on data from 16 locations over the time span covering 1979 till 2013. The results are presented in Table~\ref{tab:city_breakdown} where we provide the performance of the IFS model as the second number in each cell.

We observe that both TRU-NET CC and IFS generally achieves lower RMSE scores during the Spring and Summer months with less rainfall. By observing the Mean Error (MAE) we notice our model generally under-predicts rainfall for cities with high average rainfall (Glasgow and Cardiff) and over-predicts rainfall for cities with low average rainfall (London, Birmingham, Manchester).

\begin{table}[h!]
        \centering
        \begin{subtable}{0.49\linewidth}
        \tabcolsep=0.11cm
        \begin{tabular}{lccccc}
            \toprule
            {} &  RMSE &  R10 RMSE &  ME \\
            \midrule
            WNT & 2.155/\textbf{2.061} &     \textbf{4.019}/4.583 &         0.685/\textbf{0.075}\\
            SPR & \textbf{1.812}/2.817 &     \textbf{3.258}/8.193 &         \textbf{0.135}/0.349 \\
            SUM & \textbf{1.646}/2.884 &     \textbf{3.368}/6.805 &         \textbf{-0.095}/0.334  \\
            AUT & \textbf{1.781}/2.661 &     \textbf{2.998}/6.676 &         0.318/\textbf{0.125} \\ \hline
            All & \textbf{1.933}/2.621 &			\textbf{5.215}/6.649 & 			0.263/\textbf{0.222}\\
        \bottomrule
        \end{tabular}
        \caption{Birmingham}
    \end{subtable}
    \vfill
    \vspace*{3mm}
        \begin{subtable}{0.49\linewidth}
        \centering
        \tabcolsep=0.04cm
        \begin{tabular}{lccccc}
            \toprule
            {} &  RMSE &  R10 RMSE &  ME \\
            \midrule
            WNT & \textbf{2.292}/3.486 &     \textbf{4.737}/6.564 &         0.155/\textbf{0.009} \\
            SPR & \textbf{1.829}/3.313 &     \textbf{3.516}/6.644 &         \textbf{0.301}/0.727  \\
            SUM & \textbf{2.026}/3.735 &     \textbf{4.940}/7.769 &         -0.377/\textbf{0.345} \\
            AUT & \textbf{2.241}/3.701 &     \textbf{4.517}/8.408 &         -0.185/\textbf{0.143} \\ \hline
                All & \textbf{2.215}/3.551 &			\textbf{4.809/7.396} & 		\textbf{-0.008}/0.315  \\
            \bottomrule
        \end{tabular}
        \caption{Cardiff}
    \end{subtable}
    \hfill
        \begin{subtable}{0.49\linewidth}
        \centering
        \tabcolsep=0.04cm
        \begin{tabular}{lccccc}
            \toprule
            {} &  RMSE &  R10 RMSE &  ME  \\
            \midrule
            WNT & \textbf{3.752}/5.079 &     \textbf{7.454}/9.479 &         -1.100/\textbf{-1.025} \\
            SPR & \textbf{2.316}/3.023 &     \textbf{5.733}/7.172 &         -0.312/\textbf{-0.008} \\
            SUM & \textbf{2.455}/3.758 &     \textbf{5.244}/7.413 &         \textbf{-0.241}/0.266 \\
            AUT & \textbf{3.022}/4.128 &     \textbf{6.523}/7.812 &         \textbf{-0.460}/-0.561 \\ \hline
            All & \textbf{3.132}/4.059 & 		\textbf{1.783}/2.387 & 		-0.531/\textbf{-0.337} \\
            \bottomrule
        \end{tabular}
        \caption{Glasgow}
    \end{subtable}
    \vfill
    \vspace*{3mm}
        \begin{subtable}{0.49\linewidth}
        \centering
        \tabcolsep=0.03cm
        \begin{tabular}{lccccc}
            \toprule
            {} &  RMSE &  R10 RMSE &  ME \\
            \midrule
            WNT & \textbf{2.159}/2.611 &     \textbf{3.775}/7.768 &         0.244/\textbf{-0.031} \\
            SPR & \textbf{1.935}/2.411 &     \textbf{3.866}/6.829 &       \textbf{0.309}/0.447     \\
            SUM & \textbf{2.367}/3.156 &     \textbf{6.824}/9.787 &         \textbf{-0.040}/0.487\\
            AUT & \textbf{1.940}/2.771 &     \textbf{4.461}/9.105 &         0.083/\textbf{-0.081} \\ \hline
            All &	\textbf{2.221}/2.735 &		\textbf{7.850}/8.500 &		\textbf{0.158}/0.210\\
            \bottomrule
        \end{tabular}
        \caption{London}
    \end{subtable}
    \hfill
        \begin{subtable}{0.49\linewidth}
        \centering
        \tabcolsep=0.03cm
        \begin{tabular}{lccccc}
            \toprule
            {} &  RMSE &  R10 RMSE &  ME \\
            \midrule
            WNT & \textbf{2.372}/3.212 &     \textbf{4.252}/7.223 &         0.378/\textbf{0.271}   \\
            SPR & \textbf{2.048}/3.342 &     \textbf{4.189}/7.959 &        \textbf{0.032}/0.585      \\
            SUM & \textbf{1.998}/3.795 &     \textbf{5.186}/9.244 &         \textbf{-0.179}/0.618 \\
            AUT & \textbf{2.253}/3.428 &     \textbf{5.008}/7.783 &         -0.221/\textbf{-0.093}       \\  \hline
            All & \textbf{2.374}/3.428 	&		\textbf{6.517}/7.994 &		\textbf{0.013}/0.353 \\
            \bottomrule
        \end{tabular}
        \caption{Manchester}
    \end{subtable}
    \caption{ \textbf{Seasonally dis-aggregated Performance Metrics for TRU-NET CC: } This TRU-NET CC model was trained on data between 1979 and 2013 and tested on the $7.1 {\times} 10^3$ $\mathrm{km}^2$ region around 5 cities between 2014 and August 2019. The first/second number in each cell is the associated performance of TRU-NET/IFS.  The left hand column contains the following abbreviations for Seasons: Winter=WNT, Spring=SPR, Summer=SUM, Autumn=AUT. The units for all metrics are mm/day. The scores in bold represent the best predictive performance for each metric for each seasons. We observe that both TRU-NET CC generally achieves lower RMSE scores during the Spring and Summer months with less rainfall, and under/over predicts rainfall for cities with high/low average rainfall.} 
    \label{tab:city_breakdown}
\end{table}

\subsubsection{Distribution of Predictions}
\label{sec:distribution_of_preds}
\begin{figure}[htbp]
    \centering
    \begin{subfigure}{0.70\linewidth}
        \includegraphics[width=1\linewidth]{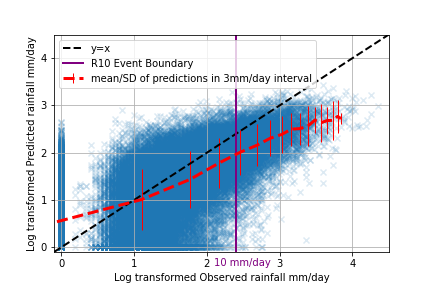}  
        \caption{TRU-NET CC}
    \end{subfigure}
    \begin{subfigure}{0.70\linewidth}
        \includegraphics[width=1\linewidth]{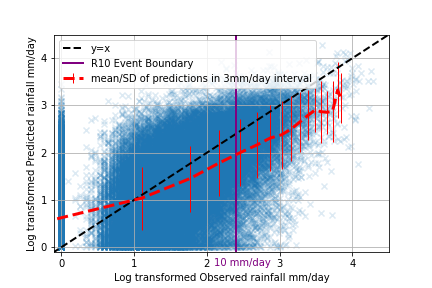}
        \caption{IFS}
    \end{subfigure}
    \caption{\textbf{Distribution of Predictions: }These figures illustrate the distribution of predicted rainfall against observed rainfall for the TRU-NET CC and IFS models from Section~\ref{sec:seasonal_breakdown}. The dashed red line shows the mean and standard deviation of predictions in 3 mm/day intervals of observed rainfall. The purple line indicates the boundary for rainfall events with at least 10mm/day. 
    For illustrative purposes, we sub-sample every 25th pair of prediction and observed value.
    The log transform used is $\operatorname{log}(y+1)$. }
    \label{fig:preds_vs_obvs}
\end{figure}

Figure~\ref{fig:preds_vs_obvs} illustrates TRU-NET CC's and IFS predicted rainfall values, for the whole UK, plotted against the true observed rainfall over the period 2014-2019. 


When comparing TRU-NET's predictions to IFS predictions, we notice a significant number of cases wherein both TRU-NET and IFS predict rainfall higher than 0mm, for days where observed rainfall is 0mm. However, as can be seen by the vertical blue cloud of points to the left of each sub-figure, TRU-NET's log-transformed predictions for non-rainy days spread up to 2.75, while IFS performs worse and spread up to 3.4.

For observed rainfall events between 10 and 19 mm/day we notice that both TRU-NET and IFS slightly under-predict the observed rainfall by a similar amount. However, TRU-NET's predictions have less variance than the IFS predictions, which routinely produce predictions significantly below or above the y=x line. This is highlighted by the large vertical spread of IFS predictions, in Figure~\ref{fig:preds_vs_obvs} (b), between observed rainfall of 10mm/day and 3.

For observed rainfall events above 20mm/day, we notice that TRU-NET under-predicts rainfall events more than IFS. We believe that the rarity of rainfall>20 events in the training set has negatively impacted TRU-NET's ability to learn these relationships, while IFS learns the underlying physical equations. 

\subsubsection{Cross Correlation across Predictions for Cities}
\label{sec:cross_correlation}

Here, we  check the spatial structure of the predictions via cross-correlation plots for TRU-NET CC Normal predictions on the central point within pairs of cities. We use Leeds as our base location and compute pairwise cross-correlations with the following six locations; Bradford (13km), Manchester (57km), Liverpool (104km), Edinburgh (261km), London (273km) and Cardiff (280km), where the each bracketed number is the distance of this location from Leeds. The cross correlations with comparison cities are ordered with increasing distance from Leeds. Linear de-trending was used.

\begin{figure}[htbp]
\begin{subfigure}{0.5\linewidth}
  \centering
  \includegraphics[trim={10 3 40 30},clip,width=1\linewidth]{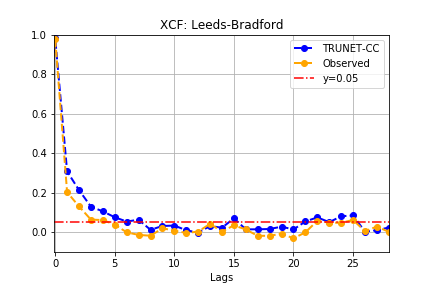}  
    \caption{XCF: Leeds-Bradford}
\end{subfigure}
\hfill
\begin{subfigure}{0.5\linewidth}
  \centering
  \includegraphics[trim={10 3 40 30},clip,width=1\linewidth]{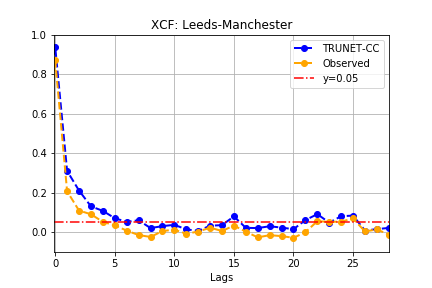}
  \caption{XCF: Leeds-Manchester}
\end{subfigure}
\vfill
\vspace*{3mm}
\begin{subfigure}{0.5\linewidth}
  \centering
  \includegraphics[trim={10 3 40 30},clip,width=1\linewidth]{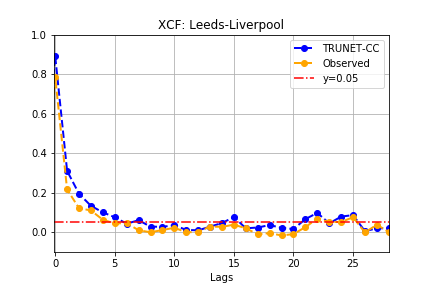} 
  \caption{XCF: Leeds-Liverpool}
\end{subfigure}
\hfill
\begin{subfigure}{0.5\linewidth}
  \centering
  \includegraphics[trim={10 3 40 30},clip,width=1\linewidth]{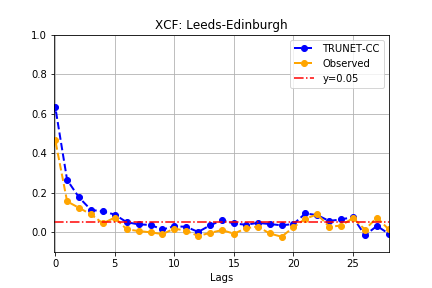} 
    \caption{XCF: Leeds-Edinburgh}
\end{subfigure}
\vfill
\vspace*{3mm}
\begin{subfigure}{0.5\linewidth}
  \centering
  \includegraphics[trim={10 3 40 30},clip,width=1\linewidth]{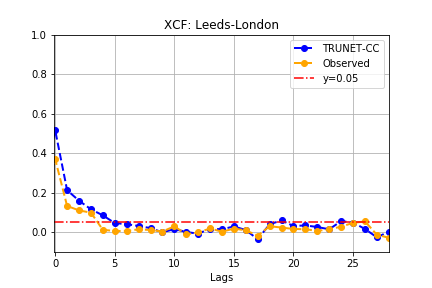} 
 \caption{XCF: Leeds-London}
 \end{subfigure}
\hfill
\begin{subfigure}{0.5\linewidth}
  \centering
  \includegraphics[trim={10 3 40 30},clip,width=1\linewidth]{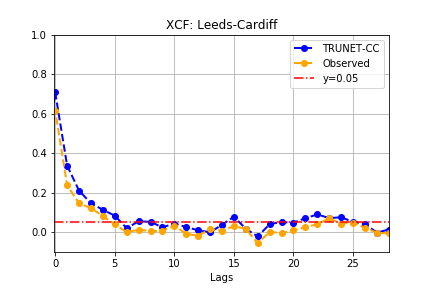} 
 \caption{XCF: Leeds-Cardiff}
 \end{subfigure}
 \caption{\textbf{Cross Correlation across Predictions: }These figures illustrate the Cross Correlation function (XCF) between rainfall predictions for Leeds and alternative cities. Here, we present XCF up to lag 28. The orange line provides the same statistics, except with the true observed rainfall values. The red line provides the 5\% significance threshold, above which we can assume their is significant correlation.}
 \label{fig:XCF}
\end{figure}

\begin{figure}[htbp]
    \centering
    \includegraphics[trim={5 5 10 0},clip,width=0.50\linewidth]{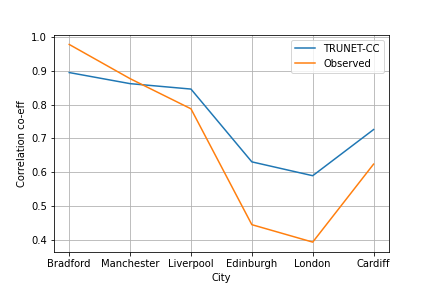}
    \caption{\textbf{Lag 0 Cross-Correlation between Leeds and other cities: }This figure shows the Cross-Correlation function (XCF) at Lag 0 for the daily rain predictions between Leeds and the 6 cities on the x-axis. The cities are ordered by increasing distance from Leeds. We use the rain prediction for the central point within the $16{\times}16$ stencil representing a city. We provide comparison to the XCF of the true observed rainfall values. }
    \label{fig:XCF_lag0}
\end{figure}

Figures~\ref{fig:XCF} and~\ref{fig:XCF_lag0} illustrate the cross correlations between TRU-NET CC's predictions for the central points of pairs of cities. Figures~\ref{fig:XCF} shows the cross-correlation function up to 28 days lag. As expected, we notice a strong correlation up to approximately 5 days. For all sets of figures, the relationships exhibited by TRU-NET CC's predictions (blue line) are approximately mirrored by the observed values (orange line) confirming that our model is producing sensible predictions. In Figure~\ref{fig:XCF_lag0}, as expected, we observe that the Lag 0 cross-correlation between the predicted daily rainfall for cities decreases as the cities become increasingly distant from each other.

\subsection{Investigation of TRU-NET's Limitations}
\label{sec:investigate_predictions}
The high temporal correlation in weather data reduces the \textit{effective sample size} and provides the risk that any neural network trained on $N$ consecutive years will only learn a limited set of weather patterns. 
The reliability of the DL model's extrapolation to out of sample predictions (new weather patterns) is more doubtful because DL models do not aim to learn the underlying physical equations, unlike numerical weather algorithms.

The three experiments introduced below evaluate the robustness of TRU-NET's out of sample predictive ability. 
\begin{figure}[htbp]
    \begin{center}
    \begin{subfigure}[!b]{0.5\linewidth}
        \includegraphics[trim={0 0 0 0}, clip,width=\linewidth]{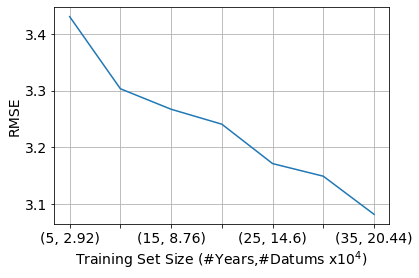}
    \caption{RMSE}
    \end{subfigure}
    \end{center}
    \vfill
    \begin{subfigure}[!b]{0.5\linewidth}\centering
        \includegraphics[trim={0 0 0 0}, clip,width=1\linewidth]{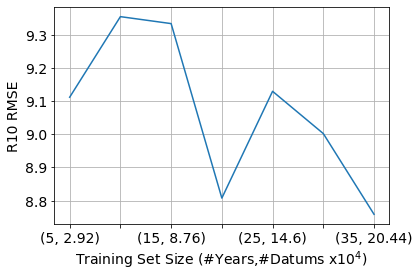}
    \caption{R10 RMSE}
    \end{subfigure}
    \begin{subfigure}[!b]{0.5\linewidth}\centering
        \includegraphics[trim={0 0 0 0}, clip,width=1\linewidth]{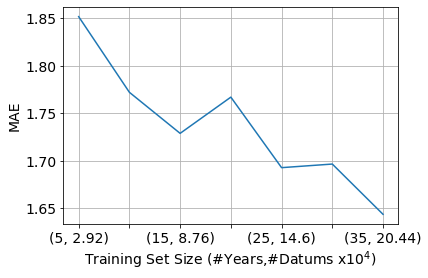}
    \caption{MAE}
    \end{subfigure}
    \caption{\textbf{Varied Time Span: } Here, we vary the size of TRU-NET CC's training set and observe the corresponding predictive performances on a test set spanning 2014 to August 2019.Sub-figures (a),(b) and (c) show the predictive performances evaluated by RMSE, R10 RMSE and MAE respectively. We observe that RMSE and MAE scores improve as the training set size increases. The units of all metrics here are mm/day. }
        \label{fig:VariedTrainingTimeSpan}
\end{figure}

\begin{figure}
    \begin{center}
    \begin{subfigure}[!b]{0.5\linewidth}
         \includegraphics[trim={10.0 7 10 10}, clip, width=\linewidth]{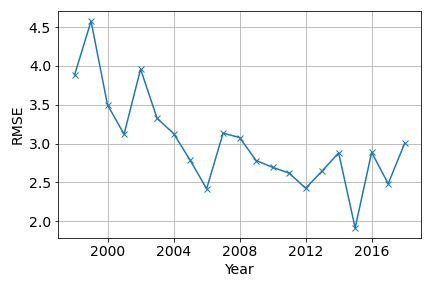}
         \caption{RMSE}
    \end{subfigure}
    \end{center}
    \vfill
    \begin{subfigure}[!b]{0.495\linewidth}\centering
         \includegraphics[trim={10.0 7 10 10}, clip,width=\linewidth]{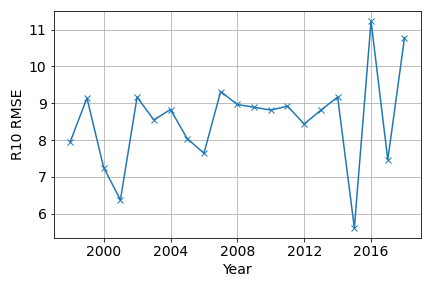}
         \caption{R10 RMSE}
    \end{subfigure}
    \begin{subfigure}[!b]{0.495\linewidth}\centering
         \includegraphics[trim={10.0 7 10 10}, clip,width=\linewidth]{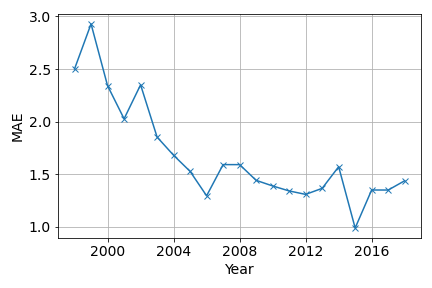}
         \caption{MAE}
    \end{subfigure}
     \caption{\textbf{Forecasting Range: }  Here, we inspect the annually aggregated predictive performance for a TRU-NET CC model trained on data spanning 1979 to 1997. We notice no clear trend in the R10 RMSE predictive performance as the test year becomes further forward in time from the training set. However, the RMSE and MAE show a steady decline between 1998 and 2005, after which point the predictive scores stay steady. The units of all metrics here are mm/day.}
    \label{fig:ForecastingRangeEvaluation}
\end{figure}

\subsubsection{Varied Time Span Experiment}
\label{sec:varied_time_span}
Here, we fix the test set to span the years 2014 to August 2019 and vary the number of years, starting from 1979, used to train our TRU-NET CC model. We measure the training set size by years and by unique test datums. As our model operates on extracted temporal patches from the coarse grid, the amount of unique datums in a training set is proportional to the product of the number of years we choose to train on and the number of locations included in the training set. In Figure~\ref{fig:VariedTrainingTimeSpan}, we observe a downward trend in RMSE and MAE as the number of years and unique test datums increases. The fact that the RMSE is reaching the lowest value for the largest dataset indicates that an increase of our dataset by using more locations could achieve further improvements in our model's predictive ability.

\subsubsection{Forecasting Range Evaluation Experiment}
\label{sec:forecast_range_evaluation}
Here, we evaluate the change in quality of predictions at increasingly larger temporal distances from the time covered by the training set. We train a TRU-NET CC model using data between 1979 and 1997 and then calculate annual RMSE, R10 RMSE and MAE metrics for each calendar year of predictions between 1998 and 2018.
In the results, illustrated in Figure~\ref{fig:ForecastingRangeEvaluation}, the R10 RMSE shows no clear upward or downward trend throughout the whole test period, while the RMSE and MAE slightly decline until 2005, after which the score remains steady. This indicates that our model's predictive ability is robust to at least 21 years of weather pattern changes due to climate change and natural decadal variability. We attribute the higher near-term RMSE and MAE errors to the increased number of R10 events between the years 1998 till 2002. Specifically, inspection of the E-obs precipitation dataset revealed that 8.3\% of the days between 1998 till 2002 experience rainfall events over 10 mm/day whereas in the period 2003 till 2007 only 6.8\% of days with rainfall events over 10 mm/day. As such there are more R10 events to contribute to the RMSE in the former period. 

\subsubsection{Varied Time Period Experiment}
\label{sec:varied_time_period_experiment}
To judge the extent to which TRU-NET's predictive ability is dependent on the time period it is trained on, we divide the 40 year dataset into 4 sub-datasets of 10 years each. The first sub-dataset (DS1) corresponds to years 1979-1988, the second sub-dataset (DS2) to the years 1989-1998, the third sub-dataset (DS3) to the years 1999-2008 and the fourth sub-dataset (DS4) to the years 2009-2018. We set-up a K-fold cross validation based experiment by training a separate TRU-NET CC model on each of DS1, DS2, DS3 and DS4, creating four models M1, M2, M3 and M4. 

Table \ref{tab:vtpe_all_results} shows the results from testing each model on the out-of-sample datasets. For each evaluation metric we perform a Tukey HSD test. A Tukey HSD test is used to accept or reject a claim that the means of two groups of values are not significantly different from each other. In our case we use the models (M1-4) as treatment groups and each models predictive scores form a groups of observations. The Tukey HSD test, then compares two models for a significant difference between the mean of each models reported predictive scores.

The Tukey HSD results for each evaluation metrics and all pairs of Models is presented in Table \ref{tab:vtpe_tukeyHSD_results}(a,b,c). The 1st two columns indicate the models under comparison, the 3rd the mean difference in their predictive scores. The rightmost column confirms whether or not there is a significant difference (sig. diff) between the performance of the corresponding pair of models. We can observe that the predictive performance between each pair of models is not significantly different. This implies our TRU-NET CC model is fairly invariant to the period of data it is trained on.

\begin{table}[htbp]
    \begin{center}
    \begin{subtable}{0.5\linewidth}\centering
        \begin{tabular}{lrrrr}
            \toprule
                                & DS1 & DS2 & DS3 & DS4 \\ \midrule
            M1 & nan              & 3.365            & 3.393            & 3.324            \\
            M2 & 3.395            & nan              & 3.385            & 3.314            \\
            M3 & 3.410            & 3.376            & nan              & 3.315            \\
            M4 & 3.420            & 3.392            & 3.400            & nan              \\ 
            \bottomrule
        \end{tabular}
        \caption{RMSE}
    \end{subtable}
    \end{center}
    \vfill 
    \begin{subtable}{0.5\linewidth}\centering
        \begin{tabular}{lrrrr}
            \toprule
                   & DS1   & DS2   & DS3   & DS4   \\ \midrule
                M1 & nan   & 9.319 & 9.399 & 9.139 \\
                M2 & 9.305 & nan   & 9.389 & 9.085 \\
                M3 & 9.410 & 9.418 & nan   & 9.210 \\
                M4 & 9.384 & 9.390 & 9.458 & nan   \\
            \bottomrule
        \end{tabular}
        \caption{R10 RMSE}
    \end{subtable}
    \begin{subtable}{0.5\linewidth}\centering
        \begin{tabular}{lrrrr}
        \toprule
        &  DS2 &  DS3 &  DS4 &  DS1 \\
        \midrule
        M1 &            1.738 &            1.794 &            1.788 &              nan \\
        M2 &              nan &            1.800 &            1.796 &            1.812 \\
        M3 &            1.766 &              nan &            1.805 &            1.835 \\
        M4 &            1.757 &            1.799 &              nan &            1.821 \\
        \bottomrule
        \end{tabular}
    \caption{MAE}
    \end{subtable}
    \caption{\textbf{Varied Time Period Predictive scores: } Here we present the predictive scores for the experiment detailed in Section~\ref{sec:varied_time_period_experiment}. We present results for RMSE (a), R10 RMSE (b) and MAE (c). We notice that M1 generally attains better predictive scores than M2, M3 and M4. }
\label{tab:vtpe_all_results}
\end{table}

\begin{table}[htbp]
    \begin{center}
    \end{center}
    \vfill
    \centering
    \begin{subtable}{\linewidth}\centering
        \begin{tabular}{ccccc}
        \toprule
        \textbf{1st model} & \textbf{2nd model} & \textbf{mean diff.} & \textbf{p-value}& \textbf{sig. diff}  \\
        \midrule
            M1 & M2 & 0.004 & 0.900    & False \\
            M1 & M3 & 0.007 & 0.900    & False \\
            M1 & M4 & 0.044 & 0.524 & False \\
            M2 & M3 & 0.003 & 0.900    & False \\
            M2 & M4 & 0.040 & 0.591 & False \\
            M3 & M4 & 0.037  & 0.635 & False \\
        \bottomrule
        \end{tabular}
        \caption{RMSE}
    \end{subtable}
    \vfill
    \vspace*{3mm}
    \begin{subtable}{\linewidth}\centering
        \begin{tabular}{ccccc}
        \toprule
        \textbf{1st model} & \textbf{2nd model} & \textbf{mean diff.} & \textbf{p-value} &  \textbf{sig. diff}  \\
        \midrule
               M1       &        M2       &      -0.026      &      0.900       &      False       \\
               M1       &        M3       &       0.060      &      0.900       &False       \\
               M1       &        M4       &       0.125      &     0.595     &False       \\
               M2       &        M3       &       0.087      &     0.796     &False       \\
               M2       &        M4       &       0.151      &     0.460     &False       \\
               M3       &        M4       &       0.065      &      0.900       &False       \\
        \bottomrule
        \end{tabular}
        \caption{R10 RMSE}
    \end{subtable}
    \vfill\vfill
    \vspace*{3mm}
    \begin{subtable}{\linewidth}\centering
        \begin{tabular}{ccccc}
        \toprule
        \textbf{1st model} & \textbf{2nd model} & \textbf{mean diff.} & \textbf{p-value} & \textbf{sig. diff}  \\
        \midrule
               M1       &        M2       &       0.029       &     0.612    &      False       \\
               M1       &        M3       &       0.029      &     0.625     &      False       \\
               M1       &        M4       &       0.019      &     0.839     &      False       \\
               M2       &        M3       &      -0.001      &      0.900       &      False       \\
               M2       &        M4       &      -0.010      &      0.900       &      False       \\
               M3       &        M4       &      -0.010      &      0.900       &      False       \\
        \bottomrule
        \end{tabular}
    \caption{MAE}
    \end{subtable}
    \caption{\textbf{Varied Time Period - Tukey HSD test: } We train four TRU-NET CC models (M 1-4) on 4 training sets, labelled (DS 1-4), which cover mutually exclusive 10 year time spans. Each model is then tested on all time spans, except that of its training set. We evaluate the predictions using three evaluation metrics (RMSE, R10 RMSE, MAE) and then for each evaluation metric perform a Tukey HSD test on the results. The final column of each table confirms that there is no statistically significant difference (sig. diff) between the mean performance of the corresponding two models. This implies that the performance of our TRU-NET CC model is invariant to the time period it is trained on, provided all the time periods have the same time length. The units of all metrics here are mm/day.}
    \label{tab:vtpe_tukeyHSD_results}
\end{table}

\subsection{Ablation: Fused Temporal Cross Attention}
\label{sec:ablation}
In this section, we investigate the efficacy of our FTCA relative to other methods for achieving the multi-scale hierarchical structure in the TRU-NET Encoder. More concretely, we replace the temporal fused cross attention with concatenation, last element method \citep{hierachicalLSTM_Lastcell,hierachical_LSTM_lasthiddenstate2} and temporal self attention \citep{HierachicalLSTM_selfattn}. We examine how the effect of changing the number of heads in FTCA.
Table~\ref{tab:ablatin} shows that our model achieves lower RMSE than other methods of achieving the multi-scale hierarchical structure. Furthermore, we notice strong performance relative to the self attention variant which has the same model size. This highlights the importance of using information from higher spatio-temporal scales to guide the aggregation of information from lower spatio-temporal scales in our TRU-NET model.

\begin{table}[htbp]
\centering
        \begin{tabular}{@{}llll@{}}
            \toprule
            TRU-NET CC  & RMSE  & R10 RMSE & MAE \\ \midrule
            T Cross-Attn 8 heads  & \textbf{3.072} & \textbf{8.729}  & \textbf{1.634} \\
            T Cross-Attn 4 heads & 3.100  & 8.985  & 1.656\\
            T Cross-Attn 1 heads & 3.100 & 8.893 & 1.656 \\
            Concatenation      & 3.147  &9.146 &1.661 \\ 
            Last Element       & 3.098  &8.861  &1.641\\
            T Self-Attn       &3.124  &   9.085   &1.646\\ 
            \bottomrule
            \end{tabular}
        \caption{\textbf{Ablation Study: } Here, we evaluate the predictive performance of alternative methods to achieve the multi-scale hierarchical structure in the TRU-NET CC Encoder. We evaluate these TRU-NET CC variants using a training set consisting of the years 1979 till 2013. The test set is composed of data between the dates 2013 till August 2019 for the whole UK. We observe that our proposed Temporal Cross Attention (T Cross-Attn) with 8 heads outperforms other methods. The units of all metrics here are mm/day. }
        \label{tab:ablatin}
\end{table}

\section{Conclusion and Future work}
\label{sec:conclusion}
In this work we present TRU-NET, designed to predict high resolution precipitation from a climate model's coarse-resolution simulation of other weather variables. TRU-NET features a novel Fused Temporal Cross Attention mechanism to improve the modelling of processes defined on multiple spatio-temporal scales. We also provide a non probabilistic adaptation of the conditional-continuous loss function to obtain predictions for zero-skewed rainfall events.

For the prediction of local precipitation over the UK, using inputs defined on a coarse resolution grid, our model achieves a 10\% lower RMSE than a hierarchical ConvGRU model and a 15\% lower RMSE than a dynamical weather forecasting model (IFS) initialised 0-24h before each precipitation prediction. After further analysis, we observe that TRU-NET attains lower RMSE scores than IFS when predicting rainfall events up to and including 20 mm/day, which comprises the majority of rainfall events. However, after this point TRU-NET under-predicts rainfall events to a higher degree than IFS.

We address concerns regarding the suitability of DL approaches to precipitation prediction \citep{rasp2020weatherbench, Rasp9684}, given the limited amount of training data. We show that the current amount of data available is sufficient for a DL approach to produce quality predictions at 10km resolution with an input of coarse-resolution variables from a numerical simulation.

The current work used deterministic models and readily available reanalysis data as an analogue for climate model output. Future works, could utilise probabilistic neural network methods, such as Monte Carlo Dropout \citep{gal2015dropout} or Horseshoe Prior\citep{ghosh2018structured}, as well as data from climate simulations to simulate risks of severe weather under varying climate scenarios. Further, methods combining Extreme Value Theory and machine learning \citep{10.1145/3292500.3330896} could be used to improve TRU-NET's ability to predict rainfall events over 20 mm/day.

\begin{acknowledgements} 
The project was funded by Alan Turing Institute under Climate Action Pilot Projects Call.  We also acknowledge support from the HPC Facilities at the University of Warwick (Orac cluster). The code and data used to train and evaluate our models can be downloaded from \href{https://github.com/Akanni96/TRUNET}{https://github.com/Akanni96/TRUNET}. We acknowledge the ERA5 \citep{ERA5, ERA5data} and the E-OBS \citep{E_obs} dataset from the Copernicus Climate Change Service, and the EU-FP6 project UERRA (\url{http://www.uerra.eu}) and the data providers in the ECA\&D project (\url{https://www.ecad.eu}). Peter Dueben gratefully acknowledges funding from the Royal Society for his University Research Fellowship and the
ESiWACE2 project which has received funding from
the European Union's Horizon 2020 research and
innovation programme under grant agreement No
823988. Peter Watson gratefully acknowledges funding (grant no.~NE/S014713/1) from the Natural Environment Research Council for his Independent Research Fellowship. Yulan He gratefully acknowledges funding (grant no.~EP/V020579/1) from the UK Research and Innovation for her Turing AI Fellowship.
We also acknowledge Dr. Sherman Lo's help in processing the datasets. 
\end{acknowledgements}

\bibliographystyle{spbasic}      

\bibliography{Bibliography}

 \end{document}